\documentstyle[prd,aps,eqsecnum,epsf]{revtex}
\def \tablerule{\noalign {\vskip3truept\hrule\vskip3truept}}
\def \up {0pt}

\def\gsim{\mathrel{
\rlap{\raise 0.511ex \hbox{$>$}}{\lower 0.511ex
\hbox{$\sim$}}}}

\def\lsim{\mathrel{
\rlap{\raise 0.511ex \hbox{$<$}}{\lower 0.511ex
\hbox{$\sim$}}}}

\begin{document}

\title{Improved filters for  gravitational waves from  
inspiralling compact binaries}

\author{Thibault Damour$^1$, Bala R. Iyer$^2$ and B.S. Sathyaprakash$^3$}

\address{$^1${\it Institut des Hautes Etudes Scientifiques, 91440
Bures-sur-Yvette, France {\rm and}  
DARC, CNRS - Observatoire de Paris, 92195 Meudon, France}}
\address{$^2${\it Raman Research Institute, Bangalore 560 080, India}}

\address{$^3${\it Cardiff University of Wales, P.O. Box 913, Cardiff,
CF2 3YB, U.K. and California Institute of Technology, Pasadena, CA 91125}}
\date{\today}
\maketitle
\begin{abstract}
The order of the post-Newtonian expansion needed, to extract
in a reliable and accurate manner the fully general relativistic
gravitational wave signal from inspiralling compact binaries,
is explored. A class of approximate wave forms, called 
$P$-approximants, is constructed based
on the following two inputs: (a) The introduction of two new 
energy-type and flux-type functions $e(v)$ and $f(v),$ respectively, 
(b) the systematic use of Pad\'e approximation for constructing 
successive approximants of $e(v)$ and $f(v)$. 
The new $P$-approximants are not only more {\em effectual} 
(larger overlaps) and more {\em faithful} (smaller biases)
than the standard Taylor approximants, but also converge faster
and monotonically. 
The presently available $O\left (\frac{v}{c} \right )^5$-accurate 
post-Newtonian results can be used to construct $P$-approximate 
wave forms that provide overlaps with the exact wave form 
larger than $96.5\%$ implying that more than 90\% of 
potential events can be detected with the aid of $P$-approximants as
opposed to a mere 10-15 \% that would be detectable using standard 
post-Newtonian approximants.
\end{abstract}
\pacs{04.3.0Db, 04.25.Nx, 04.80.Nn, 95.55.Ym}
\section {Introduction and methodology}

Inspiralling compact binaries consisting of neutron stars and/or
black holes are among the most promising candidate sources
for  interferometric detectors of gravitational
waves such as LIGO and VIRGO. The inspiral wave form enters the detector
bandwidth  during the last few minutes of evolution of the binary.
Since the wave form can, in principle, be calculated accurately,
it should be possible to track the
signal phase and hence enhance the signal-to-noise ratio by
integrating the signal for the time
during which the signal lasts in the detector band. This is achieved
by filtering the detector output with a template which
is a copy of the expected signal. Since in general relativity
the two-body problem has
not been solved the exact shape of the binary wave form is not
known and experimenters intend to use as a template
an approximate wave form computed perturbatively with the aid
of a post-Newtonian expansion
\cite{P93,CFPS,TN94,S94,TS94,BDIWW,BDI,WW,BIWW,B96,TTS97}.
Thus, template wave forms used in detection will be different from the
actual signal that may be present in the detector output. As a result the
overlap of template and signal wave forms would be less than what one would
expect if they had exactly matched.

In this paper we explore the order of the post-Newtonian expansion needed
to extract in a reliable and accurate manner the actual, fully general
relativistic signal. Previous attacks on this problem
\cite{CFPS,TN94,TTS97,Cutler,P95,P97} suggested that a very
high post-Newtonian order (maybe as high as $v^{9} /c^{9}$ beyond the
leading approximation) might be needed for a reasonably accurate signal
extraction \cite{ftn1}.  Our conclusions
are much more optimistic. We show that, starting only from the presently known
$(v/c)^5$-accurate (finite mass) post-Newtonian results
\cite{BDIWW,BDI,WW,BIWW,B96}, but using them in a novel way, we can construct new
template wave forms having overlaps larger than 96.5\% with the ``exact'' wave
forms. Since a reduction in signal-to-noise ratio by 3\% only results in a
loss in the number of events by 10\%, and since our computations indicate that
the new templates entail only small biases in the estimate of signal parameters
(see Tables \ref{effectualness.tm} and \ref {effectualness.fm} below), 
we conclude that presently known post-Newtonian results
will be adequate for many years to come.

Before entering the details of our construction, let us clarify, at the
conceptual level, the general methodology of this work. Central to our
discussion is the following data analysis problem: On the one hand, we have
some exact gravitational wave form $h^X (t;\lambda_k)$ where $\lambda_k$,
$k=1,\ldots ,n_{\lambda}$ are the parameters of the signal (comprising,
notably, the masses $m_1$ and $m_2$ of the members of the emitting
binary \cite{ftn2}).  On the other
hand, we have theoretical calculations of the motion of \cite{DD81}, and
gravitational radiation from \cite{BDIWW,BDI,WW,BIWW,B96}, 
binary systems of
compact bodies (neutron stars or black holes). The latter calculations give the
post-Newtonian expansions (expansions in powers of $v/c$) of,
essentially \cite{ftn3},
two physically important functions: an energy function $E(v)$ and a
gravitational
flux function $F(v)$ (see exact definitions below). Here, the dimensionless
argument $v$ is an invariantly defined ``velocity''
\cite{ftn4} related to the instantaneous {\it
gravitational wave}
frequency $f^{\rm GW}$ ($=$ twice the {\it orbital} frequency) by
\begin{equation}
v = (\pi \, m \, f^{\rm GW})^{\frac{1}{3}} \, , \label {eq:n1}
\end{equation}
where $m \equiv m_1 + m_2$ is the total mass of the binary. Let us denote by
$E_{T_n}$ and $F_{T_n}$ the $n^{\rm th}$-order Taylor approximants of the
energy and flux functions,
\begin{equation}
E_{T_n} = \sum_{k=0}^{n} E_k (\eta) \, v^k = E(v) + O (v^{n+1}) \, , \label
{eq:n2a}
\end{equation}
\begin{equation}
F_{T_n} = \sum_{k=0}^{n} F_k (\eta) \, v^k = F(v) + O (v^{n+1}) \, , \label
{eq:n2b}
\end{equation}
where
\begin{equation}
\eta \equiv \frac{m_1 \, m_2}{(m_1 + m_2)^2} \label {eq:n3}
\end{equation}
is the symmetric mass ratio. For finite $\eta$, the Taylor approximants
(\ref{eq:n2a}), (\ref{eq:n2b}) are known for $n \leq 5$ 
\cite{DD81,BDIWW,BDI,WW,BIWW,B96}. 
In the test mass limit, $\eta \rightarrow 0$, $E(v)$
is known exactly and $F(v)$ is known up to the order $n=11$
\cite{P93,CFPS,TN94,S94,TS94,TTS97}. [There are logarithmic terms appearing
for $n\geq 6$ that we shall duly discuss later, but in this Introduction we
simplify the notation by not introducing them.]

\begin {figure}[h]
\caption {Schematic illustration of our methodology to compute improved
templates.}
\centerline {\epsfxsize 6 true in  \epsfbox {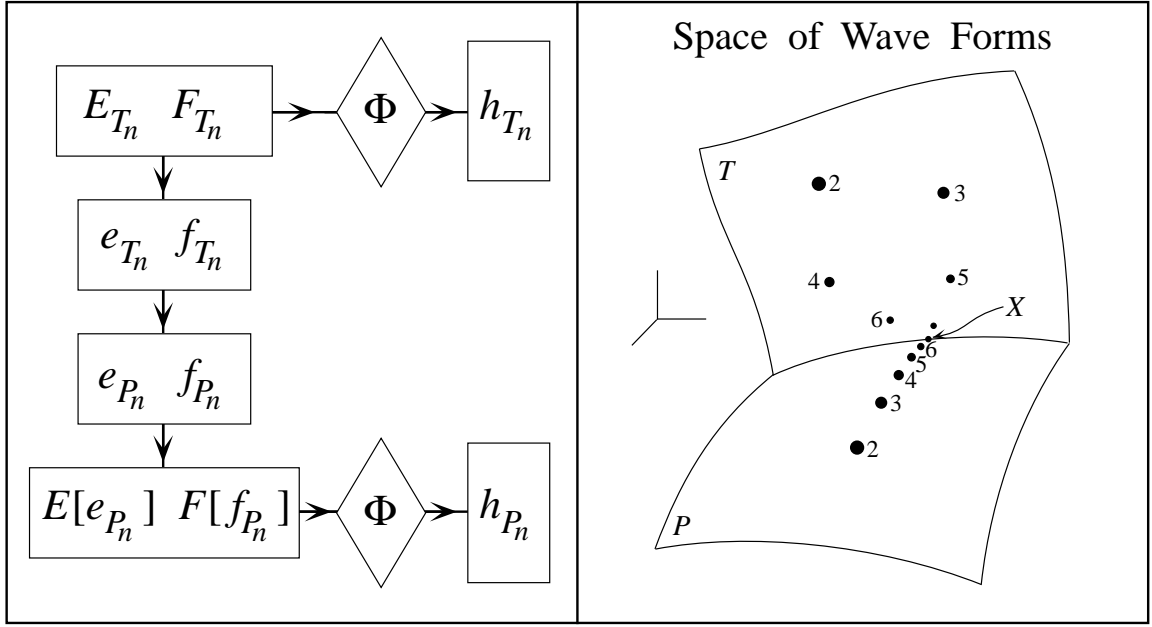} }
\label {fig:conceptual}
\end {figure}

The problem is to construct a sequence of approximate wave forms $h_n^A
(t;\lambda_k)$, starting from the post-Newtonian expansions (\ref{eq:n2a}),
(\ref{eq:n2b}). In formal terms, any such construction defines a {\it map} from
the set of the Taylor coefficients of $E$ and $F$ into the (functional) space
of wave forms (see Fig. \ref{fig:conceptual}). 
Up to now, the literature has considered only the
most standard map, say $T$,
\begin{equation}
(E_{T_n} ,F_{T_n}) \stackrel{T}{\rightarrow} h_n^T (t,\lambda_k) \, , \label
{eq:n4}
\end{equation}
obtained by inserting the successive Taylor approximants
\cite{ftn5}
 (\ref{eq:n2a}),
(\ref{eq:n2b}) into the integral giving the time evolution of the gravitational
wave phase, see e.g. \cite{Cutler,P95}. [Details are given below.] In
this work, we shall define a new map, say ``$P$'', based on a four-stage
procedure (Fig. \ref{fig:conceptual}) 
\begin{equation}
(E_{T_n} , F_{T_n}) \rightarrow (e_{T_n} , f_{T_n}) \rightarrow (e_{P_n} ,
f_{P_n}) \rightarrow (E[e_{P_n}] , F[e_{P_n} , f_{P_n}]) \rightarrow h_n^P
(t,\lambda_k) \, . \label {eq:n5}
\end{equation}

The two essential ingredients of our procedure are: (i) the introduction, on
theoretical grounds, of two new, supposedly more basic and hopefully better
behaved, energy-type and flux-type functions, say $e(v)$ and $f(v)$, and (ii)
the systematic use of Pad\'e approximants (instead of straightforward Taylor
expansions) when constructing successive approximants of the intermediate
functions $e(v)$, $f(v)$. Let us also note that we further differ from previous
attacks on the problem by using a numerical (discrete) fast Fourier transform
to compute the overlaps between the exact and approximate wave forms. We find
that the previously used analytical stationary phase approximation gives only
poor estimates of the overlaps (see Table \ref{stationary.phase}).

One of the aims of the present paper is to show that the new sequence of
templates $h_n^P (t;\lambda)$ is, in several ways, ``better'' than the standard
one $h_n^T (t;\lambda)$. In this respect, it is convenient to introduce some
terminology. We shall say that a multi-parameter family of approximate wave
forms $h^A (t;\mu_k)$, $k=1,\ldots ,n_{\mu}$ is an {\it effectual} model of
some exact wave form $h^X (t;\lambda_k)$; $k=1,\ldots ,n_{\lambda}$ (where one
allows the number of model parameters $n_{\mu}$ to be different from, i.e. in
practice, strictly smaller than $n_{\lambda}$) if the overlap, or normalized
ambiguity function, between $h^X (t;\lambda_k)$ and the time-translated family
$h^A (t-\tau ; \mu_k)$,
\begin{equation}
{\cal A} (\lambda_k ,\mu_k) = \max_{\tau,\phi} \frac{ \langle h^X
(t;\lambda),h^A (t-\tau ;\mu)\rangle}{\sqrt{\langle h^X (t;\lambda),h^X
(t;\lambda)\rangle \langle h^A (t;\mu),h^A (t;\mu)\rangle}} \, , \label {eq:n6}
\end{equation}
is, after maximization on the model parameters 
$\mu_k$ \cite {caution:phase}, larger than some given threshold, 
e.g. $\max_{\mu_k} {\cal A} (\lambda_k , \mu_k) \geq
0.965$ \cite {eventrate}. 
[In Eq. (\ref{eq:n6}) the scalar product $\langle h,g \rangle$ denotes
the usual Wiener bilinear form involving the noise
spectrum $S_n (f)$ (see below).] While an {\it effectual} model may be a
precious tool for the successful detection of a signal, it may do a poor job in
estimating the values of the signal parameters $\lambda_k$. We shall then say
that a family of approximate wave forms $h^A (t; \lambda_k^A)$, where the
$\lambda_k^A$ are now supposed to be in correspondence with (at least a subset
of) the signal parameters, is a {\it faithful} model of $h^X (t;\lambda_k)$ if
the ambiguity function ${\cal A} (\lambda_k ,\lambda_k^A)$, Eq. (\ref{eq:n6}),
is maximized for values of the model parameters $\lambda_k^A$ which differ from
the exact ones $\lambda_k$ only by acceptably small biases
\cite{ftn6}.
A necessary
\cite{ftn7} criterion for faithfulness,
and one which is very easy to implement in practice, is that the ``diagonal''
ambiguity ${\cal A} (\lambda_k ,\lambda_k^A = \lambda_k)$ be larger than, say,
0.965.

Using this terminology, we shall show in this work that our newly defined map,
Eq. (\ref{eq:n5}), defines approximants which, for practically all values of
$n$ we could test, are both more effectual (larger overlaps) and more faithful
(smaller biases) than the standard approximants Eq. (\ref{eq:n4}). A related
property of the approximants defined by Eq. (\ref{eq:n5}), is that the {\it
convergence} of the sequence $(h_n^P)_{n\in {\cal N}}$ is both faster and much
more monotonous than that of the standard sequence $(h_n^T)_{n\in {\cal N}}$. This
will be shown below in the (formal) test mass limit $\eta \rightarrow 0$ where
one knows both the exact functions $E(v)$ and (numerically) $F(v)$ 
\cite{P95},
and their Taylor expansions to order $v^{11}$ 
\cite{TTS97}. The convergence
will be studied both ``visually'' (by plotting successive approximants to $E$
and $F$) and ``metrically'' (by using the ambiguity function (\ref{eq:n6}) to
define a distance between normalized wave forms). Most of our convergence tests
utilize the rich knowledge of the post-Newtonian expansions (\ref{eq:n2a}),
(\ref{eq:n2b}) in the test mass limit $\eta \rightarrow 0$. The very
significant qualitative and quantitative advantages of the new sequence of
approximants, Eq. (\ref{eq:n5}), over the standard one, Eq. (\ref{eq:n4}), when
$\eta \rightarrow 0$, make it plausible that the new sequence $(h_n^P)$ will
also fare much better in the finite mass case $0 \not= \eta \leq \frac{1}{4}$.
This question, that we can call the problem of the {\it robustness} of our
results under the deformations brought by a finite value of $\eta$ in the
coefficients $E_k (\eta)$, $F_k (\eta)$ in Eqs. (\ref{eq:n2a}), (\ref{eq:n2b}),
is more difficult to investigate, especially because one does not know, in this
case, the ``exact'' results for $E(v;\eta)$ and $F(v;\eta)$. We could, however,
check the robustness of our construction in two different ways: (i) by studying
the ``Cauchy criterion'' for the convergence of the (short) sequence $(h_0^P
(\eta), h_2^P (\eta), h_4^P (\eta), h_5^P (\eta))$ versus that of the
corresponding Taylor sequence, and (ii) by introducing a one-parameter family
of fiducial ``exact'' functions $e_{\kappa_0}^X (v)$, $f_{\kappa_0}^X (v)$ to
model the unknown higher-order $(n \geq 6)$ $\eta$-dependent contributions to
the post-Newtonian expansions (\ref{eq:n2a}), (\ref{eq:n2b}) and by studying
for a range of values of the parameter $\kappa_0$ the convergence of the short
sequence $(h_0^P (\eta) , \ldots , h_5^P (\eta))$ toward the fiducial ``exact''
wave form $h_{\kappa_0}^X (\eta)$.

Though we believe the work presented below establishes the superiority of the
new approximants $h_n^P$ over the standard ones $h_n^T$ and shows the practical
sufficiency of the presently known $v^5$-accurate
post-Newtonian results, we still think that it is an important (and
challenging) task to improve the (finite mass) post-Newtonian results. Of
particular importance would be the computation
\cite{3pn} of the $v^6$-accurate (equations
of motion and) energy function in confirming and improving our estimate below
of the location of the last stable orbit for $\eta \not= 0$. Our calculations
also suggest that knowing $E$ and $F$ to $v^6$ would further improve the
effectualness (maximized overlap larger than 98\%) and, more importantly, the
faithfulness (diagonal overlap larger than 99.5\%) to a level allowing a loss in
the number of detectable events smaller than 1\%, and significantly smaller
biases (smaller than 0.5\%) in the parameter estimations than 
the present $O(v^5)$ results (about 1---5\%).

The rest of this paper is organized as follows: In Sec. \ref {sec:phase}
we briefly discuss the phasing of restricted post-Newtonian
gravitational wave forms, wherein corrections are only included to the phase
of the wave form and not to the amplitude, indicating the way in which
energy and flux functions enter the phasing formula. 
Various forms of energy and flux functions are introduced
in Secs. \ref {sec:energy}
and \ref {sec:flux}, respectively and their performance compared.
The ambiguity function, which
is the overlap integral of two wave forms as a function of their
parameters, is discussed in Sec. \ref{sec:ambiguity}
and some details of its computation by a numerical fast
Fourier transform are given.
In Sec.\ref{sec:results} we present
the results of our computations in the test mass case
while in Sec.\ref{sec:robustness} we 
investigate the robustness of these test mass results as completely
as possible. Sec.\ref{sec:conclusions} contains 
our summary and concluding remarks.
The paper concludes with two appendices. In Appendix~\ref {appendix:pade}
we discuss the Pad\'e approximants,  their relevant 
useful properties and list some useful formulas used in
the computations. In Appendix~\ref {appendix:phase} we discuss carefully
the issue of optimizing over the phases and provide 
a clear geometrical picture to implement the procedure.

\section {The phasing formula}\label {sec:phase}

To get an accurate expression for the evolving wave form $h_{ij} (t)$
emitted by an inspiralling compact binary one needs, in principle, to
solve two interconnected problems: (i) one must work out (taking into
account propagation and nonlinear effects) the way the material source
generates a gravitational wave, and (ii) one must simultaneously work out
the evolution of the source (taking into account radiation-reaction
effects). The first problem, which in a sense deals mainly with the
(tensorial) {\it amplitude} of the gravitational signal is presently
solved to order $v^5$ \cite{BDIWW,BDI,WW,BIWW,B96}.
Such an approximation on the instantaneous amplitude $h_{ij}$ seems quite
sufficient in view of the expected sensitivity of the LIGO/VIRGO network.
On the other hand, the second problem, which determines the evolution of
the {\it phase} of the gravitational signal, is crucial for a successful
detection. For simplicity, we shall work here within the ``restricted wave
form'' approximation\cite{ftn8},
i.e. we shall focus on the main Fourier component of the signal,
schematically $h(t) = a^{\rm GW} (t) \cos \phi^{\rm GW} (t)$, where the
gravitational wave phase $\phi^{\rm GW}$ is essentially, in the case of a
circular binary, twice the orbital phase $\Phi : \phi^{\rm GW} (t) = 2\Phi (t)$.

We find it conceptually useful to note the analogy between the radio-wave
observation of binary pulsars and the gravitational-wave observation of a
compact binary. High-precision observations of binary pulsars make a crucial use
of an accurate ``timing formula'' \cite{DD}
\begin{equation}
\phi_n^{\rm PSR} = F[t_n ; p_i] \, , \label {eq:N1}
\end{equation}
linking the rotational phase of the spinning pulsar (stroboscopically
observed when $\phi_n^{\rm PSR} = 2\pi n$ with $n\in {\cal N}$) to the time of
arrival $t_n$ on Earth of an electromagnetic pulse, and to some parameters
$p_i$. Similarly, precise observations of an inspiralling compact binary,
will need an accurate ``phasing formula'', i.e. an accurate mathematical
model of the continuous evolution of the gravitational wave phase
\begin{equation}
\phi^{\rm GW} = 2\Phi = F[t;p_i] \, , \label {eq:N2}
\end{equation}
involving a set of parameters $\{ p_i \}$ carrying information about the
emitting binary system (such as the two masses $m_1$ and $m_2$).

Heuristically relying on a standard energy-balance argument, the time
evolution of the orbital phase $\Phi$ is determined by two functions: an
energy function $E (v)$, and a flux function $F(v)$. Here the
argument $v$ is defined by Eq. (\ref{eq:n1}) which can be rewritten in terms of
the instantaneous {\it orbital} angular frequency $\Omega$
\begin{equation}
v \equiv (m\Omega)^{1/3} \equiv x^{1/2} \label {eq:N3}
\end{equation}
(as above $m \equiv m_1 + m_2$ denotes the total mass of the binary). The
(dimensionless) energy function $E$ is defined by
\begin{equation}
E_{\rm tot} = m (1+ E ) \label {eq:N4}
\end{equation}
where $E_{\rm tot}$ denotes the total relativistic energy (Bondi mass) of
the binary system. The flux function $F(v)$ denotes the gravitational
luminosity of the system (at the retarded instant where its angular
velocity $\Omega$ is given by Eq. (\ref{eq:N3})). Note that the three
quantities $v$, $E$ and $F$ are invariantly defined (as global
quantities in the instantaneous center of mass frame), so that the two
functions $E (v)$, $F(v)$ are coordinate-independent constructs.
Denoting as above the symmetric mass ratio by $\eta \equiv m_1 m_2 / (m_1 +
m_2)^2$, the energy balance equation $dE_{\rm tot} / dt =-F$ gives the following
parametric representation of the phasing formula Eq. (\ref{eq:N2}) 
(written here for the orbital phase)
\begin{equation}
t(v) = t_c +m \int_v^{v_{\rm lso}} dv \, \frac{{E}'(v)}{F(v)} \, , \label{eq:6a}
\end{equation}
\begin{equation}
\Phi (v) = \Phi_c + \int_v^{v_{\rm lso}} dv v^3 \, \frac{{E}'(v)}{F(v)} \, ,
\label {eq:6b}
\end{equation}
where $t_c$ and $\Phi_c$ are integration constants, and where for lisibility we
have not introduced a new name (such as $v'$) for the dummy integration
variable. Note that $E' (v) < 0$, $F(v) > 0$ so that both $t$ and $\Phi$
increase with $v$. For definiteness, we have written the integrals in Eqs. 
(\ref{eq:6a}), (\ref{eq:6b}) in terms of a specific reference velocity, chosen
here to be the velocity corresponding to the last stable circular orbit of the
binary. Note that the choice of such a reference point is, in fact, entirely
arbitrary and a matter of convention as one introduces the two integration
constants $t_c$ and $\Phi_c$ (which will be optimized later). The choice
$v_{\rm ref} = v_{\rm lso}$ is technically and physically natural as it is the
value where the integrand vanishes (because of $E'(v)$). The definition (and
properties) of our approximants do not depend on this choice and the reader is
free to use instead his/her favorite reference point. On the other hand, what is not a
matter of convention is that, in absence of information about the coalescence
process, we shall also use $v_{\rm lso}$ to define the time when the
inspiral wave
form shuts off.

The numerical value of $v_{\rm lso}$ in the case of
a test mass orbiting a black hole (i.e. the limiting case $\eta
\rightarrow 0$) is $1/\sqrt {6}$. In the case of binaries of comparable
masses $(\eta \not= 0)$ $v_{\rm lso}$ is the value of $v$ where ${E}'
(v)$ vanishes. We will discuss below ways of estimating $v_{\rm lso}
(\eta)$. Knowledge of $v_{\rm lso}$ (considered now has a physical quantity
affecting the signal and not as a simple reference point) is important in
gleaning astrophysical information since the inspiral wave form would shut off
at that point and the coalescence wave form, whose shape depends on equation of
state of stars, etc.,  would begin. One of the questions we address below is
whether (as had been suggested \cite{P95}) knowledge of $v_{\rm lso} (\eta)$
is crucial for getting accurate inspiralling wave templates.

To warm up, let us recall that in the ``Newtonian'' approximation (i.e.
when using the quadrupole formula for the gravitational wave emission) one
has
\begin{equation}
{E}_N (v) = -\frac{1}{2} \eta v^2 \ ; \ F_N = \frac{32}{5} \eta^2
v^{10} \, , \label {eq:N6}
\end{equation}
so that the above formulas reduce (after redefining the constants 
of integration, or, equivalently, formally setting $v_{\rm lso} = \infty$) to 
\begin{equation}
t = t_c - \frac{5}{256} \, m \, \eta^{-1} \, v^{-8} \;;\;
 \Phi = \Phi_c - \frac{1}{32} \,
\eta^{-1} \, v^{-5} \, . \label{eq:9}
\end{equation}
The explicit Newtonian phasing formula is obtained by eliminating
$v$ and given by
\begin{equation}
\Phi_N (t) = \Phi_c -\left(\frac{t_c-t}{5\tau}\right )^{\frac{5}{8}} \ ; \
\hbox{where} \ \tau \equiv \eta^{3/5} \, m \ (\hbox{``chirp time scale''}).
\label{eq:10}
\end{equation}

The corresponding Newtonian gravitational wave amplitude is (for some constant
$C$)
\begin{equation}
a_N^{\rm GW} (v) = C \, v^2 \, , \label{eq:n7}
\end{equation}
so that the explicit Newtonian templates read
\begin{equation}
h_N (t) = C' (t_c -t)^{-\frac{1}{4}} \, \cos \left( 2\Phi_c - 2 \left(
\frac{t_c -t}{5\tau}\right)^{\frac{5}{8}} \right) \, . \label{eq:n8}
\end{equation}

The crucial issue for working beyond the Newtonian approximation is the
availability of sufficiently accurate representations  for the two functions
${E}' (v)$ and $F(v)$. In the astrophysically
interesting case of two comparable masses orbiting around
each other neither of the functions $E (v)$ or $F(v)$ is known
exactly and thus one must rely on a post-Newtonian expansion for both these
quantities. The question is how accurate should our knowledge of the
`energy function' $E(v)$ and the  `flux function' $F(v)$ be so that we have only
an acceptable reduction in the event rate and a tolerable bias in the estimation
of parameters. Given some approximants of the energy and flux functions
(as functions of $v$), say $E_A (v)$, $F_A (v)$, and given some fiducial
velocity \cite{ftn9} $v_{\rm lso}^A$, we shall
define a corresponding approximate template
\begin{equation}
h^A = h^A (t;C,t_c ,\Phi_c ,m,\eta) \, , \label{eq:n9}
\end{equation}
by the following parametric representation in terms of $v$:
\begin{equation}
h^A (v) = C \, v^2 \, \cos \, 2\Phi_A (v) \, , \label{eq:n10a}
\end{equation}
\begin{equation}
t(v) = t_c +m \int_v^{v_{\rm lso}^A} dv \, \frac{E'_A (v)}{F_A(v)} \, ,
\label{eq:n10b}
\end{equation}
\begin{equation}
\Phi_A (v) = \Phi_c + \int_v^{v_{\rm lso}^A} dv v^3 \, \frac{E'_A (v)}{F_A
(v)} \, .
\label {eq:n10c}
\end{equation}
To compute explicitly $h^A (t)$ we numerically invert Eq. (\ref{eq:n10b}) to get
$v = V_A (t)$ and substitute the result in the other equations: $h^A (t) = C \,
V_A^2 (t) \, \cos \, [2\Phi_A (V_A (t))]$. Note that we use the Newtonian
approximation for the amplitude as a function of $v$. We could use a more
refined
approximation, such as an effective (main Fourier mode) scalar amplitude
$a_A^{\rm GW} (v) \propto \Omega^{-1} \, F_A^{1/2} \propto v^{-3} \,
F_A^{1/2} (v)$. However, our main purpose here being to study the influence
of the choice of better approximants to the phase evolution on the quality
of the
overlaps, it is conceptually cleaner to stick to one common approximation
for the
amplitude (considered as a function of our principal independent variable, $v$).

The standard approximants for $E(v)$ and $F(v)$ are simply to use their
successive Taylor approximants, Eqs. (\ref{eq:n2a}), (\ref{eq:n2b}). Our
strategy
for constructing new approximants to $E(v)$ and $F(v)$ is going to be
two-pronged.
On the one hand, using the knowledge of these functions in the test-mass
limit and
general theoretical information about their mathematical structure we shall
motivate the use of representations of $E(v)$ and $F(v)$ based on other,
supposedly more basic energy-type and flux-type functions, say $e(v)$ and
$f(v)$.
On the other hand, we shall construct Pad\'e-type approximants, say $e_{P_n}$,
$f_{P_n}$, for the ``basic'' functions $e(v)$, $f(v)$, instead of
straightforward
Taylor approximants. We shall then compare the performance of the various
phasing
formulas defined by inserting in Eqs. (\ref{eq:n10a})--(\ref{eq:n10c})
either the
standard $E_A^{\rm old} = E_{T_n}$, $F_A^{\rm old} = F_{T_n}$, Eqs.
(\ref{eq:n2a}), (\ref{eq:n2b}), or the new, two-stage approximants
$E_A^{\rm new}
= E[e_{P_n}]$, $F_A^{\rm new} = F[e_{P_n} , f_{P_n}]$. [In all cases, the
approximant of the derivative $E'_A (v)$ is just $d \, E_A (v) / dv$.]

\section{Energy Function} \label {sec:energy}

Let us motivate the introduction of a new energy function $e(v)$ as a more basic
object, hopefully better behaved than the total relativistic mass-energy $E_{\rm
tot}$, Eq. (\ref{eq:N4}), of the binary system. For this, let us consider the
limit $m_2 / m_1 \rightarrow 0$. In this test body limit, {\it i.e} a test
particle
$m_2$ moving in the background of a Schwarzschild black hole of mass $m_1$, the
total conserved mass-energy of the binary system reads
\begin{equation}
E_{\rm tot} = m_1 + {\cal E}_2 = m_1 - k_{\mu} p_2^{\mu} \, ,
\label{eq:N7}
\end{equation}
where $k_{\mu}$ is the time-translation Killing vector, and $p_2^{\mu}$
the 4-momentum of the test mass. [The quantity ${\cal E}_2 \equiv -k_{\mu}
p_2^{\mu}$ is the well-known conserved relativistic energy of a test
particle moving in a stationary background.] At infinity $k^{\mu} =
p_1^{\mu} / m_1$, so that the formal expression of $E_{\rm tot}$ is
$E_{\rm tot} = m_1 -(p_1 \cdot p_2)/m_1$. This expression is clearly very
asymmetric in the labels 1 and 2 and has bad analytical properties as a function
of $m_1$. Both problems are cured by working instead with the standard
Mandelstam
variable $s=E_{\rm tot}^2 = -(p_1 + p_2)^2 = m_1^2 + m_2^2 - 2 (p_1 \cdot p_2)$.
Further, it is known that, in quantum two-body problems, the symmetric
quantity
\begin{equation}
\epsilon \equiv \frac{s-m_1^2 -m_2^2}{2m_1 m_2} \equiv \frac{E_{\rm tot}^2 -
m_1^2 -m_2^2}{2m_1 m_2} \, , \label{eq:nn9}
\end{equation}
is the best energy function to consider when trying to extend
one-body-in-external-field results to two-body results 
\cite{IZ}. In the limit
$m_2 \ll m_1$ the quantity $\epsilon$ reduces simply to 
$\epsilon = -(p_1 \cdot p_2) / m_1 m_2 = {\cal E}_2 / m_2 + O (\eta)$.

In the case of a test mass in circular orbit around a Schwarzschild black hole
the explicit expression of the quantity $\epsilon$ in terms of the
invariant argument $x=v^2 \equiv (m\Omega)^{2/3}$, Eq. (\ref{eq:N3}), is
\begin{equation}
\epsilon=\frac{1-2x}{\sqrt{1-3x}}.
\label{eq:ETM}
\end{equation}
The explicit test-mass result (\ref{eq:ETM}) suggests that the (unknown)
exact two-body function $\epsilon (x)$ will have also some $\sim
(x-x_0)^{-1/2}$ singularity in the complex $x$-plane. This lead us finally to
consider, instead of the function $\epsilon$, its square or, equivalently,
the new energy function
\begin{equation}
e(x) \equiv \epsilon^2 -1 \equiv \left( \frac{E_{\rm tot}^2 - m_1^2
-m_2^2}{2m_1 m_2}\right)^2 -1 \, . \label{eq:N9}
\end{equation}
Note that we assume here that the total instantaneous relativistic energy of a
binary system (in the center of mass frame) can be defined as a time-symmetric
functional of positions and velocities (so that $E(v)$ depends on $v$ only
through
$x \equiv v^2$), as the quantity ${\tilde E}^{\rm even}$ discussed in Sec.
VII of
Ref. \cite{BD88}. It remains, however, unclear whether such a quantity is well
defined at very high post-Newtonian orders and whether it is then related to the
gravitational wave flux by the standard balance equation.

Summarizing, our proposal is to use as basic (symmetric) energy function
the quantity $e(x)$, Eq. (\ref{eq:N9}), instead of $E(x) \equiv (E_{\rm
tot} - m_1
- m_2) / (m_1 + m_2)$. Given any (approximate or fiducially ``exact'') function
$e(x)$, we shall then define the corresponding function $E (x)$ (with $x \equiv
v^2$) entering the phasing formulas (2) by solving Eq. (\ref{eq:N9}) in terms of
$E_{\rm tot} \equiv (m_1 + m_2)(1+ E )$. Explicitly, this gives
\begin {equation}
E(x) = \left [1 + 2 \eta
\left (\sqrt {1+e(x)} - 1\right ) \right ]^{1/2} - 1.
\label{eq:Etilde1}
\end{equation}
The associated $v$-derivative entering the phasing formula reads
\begin {equation}
E'(v) = 2 v\left .\frac {d E(x)}{dx} \right |_{x=v^2}=
\left .\frac {v\eta}{\left (1+ E(x)\right) \sqrt {1+e(x)}}
\frac {de(x)}{dx}
\right |_{x=v^2}.
\label {eq:etildeprime}
\end{equation}
Having defined our new, basic energy function $e(x)$, it remains to define the
approximants of $e(x)$ that we propose to use, when one knows only the Taylor
expansion of $E(x)$. For guidance, let us note that by inserting Eq.
(\ref{eq:ETM}) into Eq. (\ref{eq:N9}) one gets the following exact
expression for
the test-mass limit of the function $e(x)$
\begin{equation}
e(x;\eta =0)=-x \frac{1-4x}{1-3x}=-x(1-x-3x^2-9x^3-\cdots
-3^{n-1}x^n-\cdots). \label{eq:ETMT}
\end{equation}

The generalization of the expansion Eq. (\ref{eq:ETMT}) to non zero values of
$\eta$ is only known to second post-Newtonian (2PN) accuracy. Using Eq.
(4.25) of Ref. \cite{BDI}, that is,
\begin{equation}
E_{2PN} (x) = -\frac{1}{2} \, \eta \, x \, \left[ 1-\frac{1}{12} \,
(9+\eta) \, x
-\frac{1}{8} \, \left( 27 - 19\eta + \frac{\eta^2}{3} \right) \, x^2
\right] \, ,
\label{eq:n18}
\end{equation}
we compute the 2PN expansion of the function $e(x)$ for a finite $\eta$:
\begin{equation}
e_{\rm 2PN}
(x;\eta) =-x\left[1-(1+\frac{\eta}{3})x-(3-\frac{35}{12}\eta)x^2\right] \, .
\label{eq:18}
\end{equation}
The basic idea behind our proposal is that on the grounds of mathematical
continuity\cite{ftn10} between the case $\eta
\rightarrow 0$ and the case of finite $\eta$ one can plausibly expect the exact
function $e(x)$ to be meromorphically extendable in at least part of the complex
plane and to admit a simple pole singularity on the real axis $\propto (x-x_{\rm
pole})^{-1}$ as nearest singularity in the complex $x$-plane. We do not know the
location of this singularity when $\eta \not= 0$, but {\it Pad\'e
approximants} are
excellent tools for giving accurate representations of functions having
such pole singularities. For example, if we knew only the 2PN-accurate
(i.e. $O (v^4)$) expansion of the test-mass energy function (6), namely
$e_{\rm 2PN} (x;\eta =0)=-x (1-x-3x^2)$, its corresponding $v^4$-accurate
diagonal
Pad\'e approximant would be uniquely defined 
(see Appendix~\ref{appendix:pade}) as
\begin{equation}
e_{P_4} (x;\eta =0) =-x \frac{1-4x}{1-3x} \, , \label{eq:N10}
\end{equation}
which coincides with the {\it exact} result, Eq. (\ref{eq:ETMT}). Having
reconstructed the exact function $e(x)$, we have also reconstructed, using
only the information contained in the 2PN-accurate expansion, the existence
and location of a last stable orbit. Indeed, using Eqs. 
(\ref{eq:etildeprime}) and  (\ref{eq:N10}) we find
\begin{equation}
{E}'_{P_4} (v) = -\eta v \frac{1-6v^2}{(1-3v^2)^{3/2}} \, ,
\label{eq:N11}
\end{equation}
which is the exact test mass expression exhibiting a last stable orbit at
$v_{\rm lso} =1/\sqrt 6$. In Table~\ref{lso.tm} we have compared at different
post-Newtonian orders the $x_{\rm lso}\equiv v^2{\rm lso}$ predicted by the 
standard post-Newtonian series and the Pad\'e approximation to the same.

It is important to note that our assumption of structural stability between
$e(x;\eta =0)$ and $e(x;\eta)$ with $0 < \eta \leq \frac{1}{4}$ is internally
consistent in the sense that the coefficients of $x$ and $x^2$ in the square
brackets of Eq. (\ref{eq:18}) {\it fractionally} change, when $\eta$ is
turned on,
only by rather small amounts: $\eta / 3 \leq \frac{1}{12} \simeq +8.3\%$
and $-35
\eta / 36 \leq -35/144 \simeq -24.3\%$, respectively. This contrasts with other
attempts to consider $\eta$ as a perturbation parameter, such as Ref.
\cite{KWW}.
Indeed, in the quantities considered in the latter work several of the
$2PN$ terms
have coefficients that vary by very large fractional amounts as $\eta$ is turned
on: some examples being $12+29 \eta$, $2+25 \eta +2\eta^2$, $4+41 \eta +
8\eta^2$
in Eqs. (2.2) of the second reference in \cite{KWW}. Moreover, the fact
that many
of the coefficients in their Eqs. (2.2) {\it increase} when $\eta$ is turned on
(like the ones quoted above) is not a good sign for the reliability of their
approach as it means, roughly, that the radius of convergence of the particular
series they consider tends to {\it decrease} as $\eta$ is turned on. We shall
attempt below to further test the robustness of our proposal.

In summary, our proposal is the following: Given some usual Taylor
approximant to
the normal energy function, $E_{T_{2n}} = -\frac{1}{2} \, \eta \, x \,
(1+E_1 \, x
+ E_2 \, x^2 + \cdots + E_n \, x^n)$, one first computes the corresponding
Taylor
approximant for the $e$ function\cite{ftn11}, say
\begin {equation}
e_{{\rm T}_{2n}} =  -x \sum_{k=0}^n a_k x^k \, , \label{eq:23}
\end {equation}
in which the only known coefficients are
\begin {equation}
a_0 = 1;\ \ a_1 = -1 - \frac{\eta}{3};\ \ a_2 = -3+ \frac {35\eta}{12} \, .
\label {as}
\end {equation}
Then, one defines the improved approximant corresponding to Eq. (\ref{eq:23}) by
taking the diagonal ($P_m^m$, if $n=2m$) or subdiagonal ($P_{m+1}^m$, if
$n=2m+1$)
Pad\'e approximant of $-x^{-1} \, e_{T_{2n}} \, (x)$:
\begin{equation}
e_{P_{2n}} \, (x) \equiv -x \, P_{m+\epsilon}^m \, \left[ \sum_{k=0}^{n} a_k \,
x^k \right] \, , \label{eq:n25}
\end{equation}
where $\epsilon =0$ or 1 depending on whether $n \equiv 2m+\epsilon$ is even or
odd. For completeness, we recall the definition and basic properties of Pad\'e
approximant in Appendix~\ref{appendix:pade}. 
Let us only mention here that the $P_{m+\epsilon}^m$
approximants are conveniently obtained as a continued fraction.
For instance, the Pad\'e approximant of the $2PN$-approximate $e_{\rm T_4} (x) =
-x(a_0 + a_1 x+a_2 x^2)$ is:
\begin {equation}
e_{\rm P_4}(x) = \frac {-x c_0}{1+\frac {c_1 x}{1+c_2 x}}=
\frac {-c_0 x \left (1+c_2 x\right )} {1+\left (c_1+c_2 \right) x}.
\label{eq:ecf4}
\end{equation}
By demanding that this agrees with $e_{\rm T_4}$
to order $v^4$ we can relate the $c_n$'s in the above
equation to the $a_n$'s in Eq. (\ref{as}):
\begin {equation}
c_0 = a_0 ;\ \
c_1 = -\frac {a_1}{a_0};\ \ {\rm and}\ \
c_2 = -\frac {a_2}{a_1} + \frac {a_1}{a_0}.\ \
\label {eq:cns}
\end {equation}
Explicitly, this gives,
\begin{equation}
c_0 =1\ ; c_1 = 1+\frac{\eta}{3} \ ; \ c_2 = -c_1 - \frac{3-\frac{35}{12} \, \eta}{1 +
\frac{1}{3} \, \eta} = -\frac{4-\frac{9}{4} \, \eta + \frac{1}{9} \,
\eta^2}{1+\frac{1}{3} \, \eta} \, , \label{eq:n28}
\end{equation}
so that
\begin{equation}
e_{P_4} (x) = -x \, \frac{1+\frac{1}{3} \, \eta - \left( 4-\frac{9}{4} \, \eta +
\frac{1}{9} \, \eta^2 \right) x}{1+\frac{1}{3} \, \eta - \left(
3-\frac{35}{12} \,
\eta \right) x} \, . \label{eq:n29}
\end{equation}
Given a continued fraction approximant $e_{{\rm P}_n}(x)$ of the
truncated Taylor series $e_{{\rm T}_n}$ of the energy function $e(x)$
the corresponding $E(x)$ and $E'(x)$ functions are obtained using:
\begin {equation}
E_{{\rm P}_n}(x) = \left [1 + 2 \eta
\left (\sqrt {1+e_{{\rm P}_n}(x)} - 1\right ) \right ]^{1/2} - 1,
\label{eq:Etildecf}
\end{equation}
\begin {equation}
E_{{\rm P}_n}'(v) = 2v \left. \frac {d E_{{\rm
P}_n}(x)}{dx}\right|_{x=v^2} = \left. \frac {v\eta}{\left (1+ E_{{\rm
P}_n}(x)\right)  \sqrt {1+e_{{\rm P}_n}(x)}} \frac {de_{{\rm
P}_n}(x)}{dx}\right|_{x=v^2}. \label{eq:Etildecfprime}
\end{equation}
Thus, for instance
\begin {equation}
{\widehat E}'_{P_4} (v) \equiv \frac {- E_{\rm P_4}'(v)} {\eta v}=
\frac{c_0 (1 + 2 c_2 v^2 + (c_1 c_2 + c_2^2) v^4)}
{\left (1 + (c_1 + c_2) v^2 \right)^2
\left [1+ E_{\rm P_4}(v^2)\right] \sqrt{1 + e_{\rm P_4}(v^2)}},
\label{eq:NN1}
\end{equation}
where $E_{{\rm P}_n}$ is given by Eq. (\ref{eq:Etildecf}). The hatted notation
introduced in the left-hand side of Eq. (\ref{eq:NN1}) will again be used below
and indicates that one is dividing some function of $v$ by its Newtonian
approximation: e.g. ${\widehat E}' (v) \equiv E'(v) / E'_N (v)$ where, from Eq.
(\ref{eq:N6}) $E'_N (v) = -\eta \, v$.

\begin {figure}[h]
\caption {Exact energy functions (a) $\hat e(v)$ and
(b) $\hat E'(v),$  in the test mass case and 
$T$- and $P$-approximants in the comparable mass 
(with $\eta=1/4$) case. Note that the comparable mass case 
$T$-approximant and $P$-approximant, are smooth deformations of the 
test mass function.}
\centerline {\epsfxsize 6 true in  \epsfbox {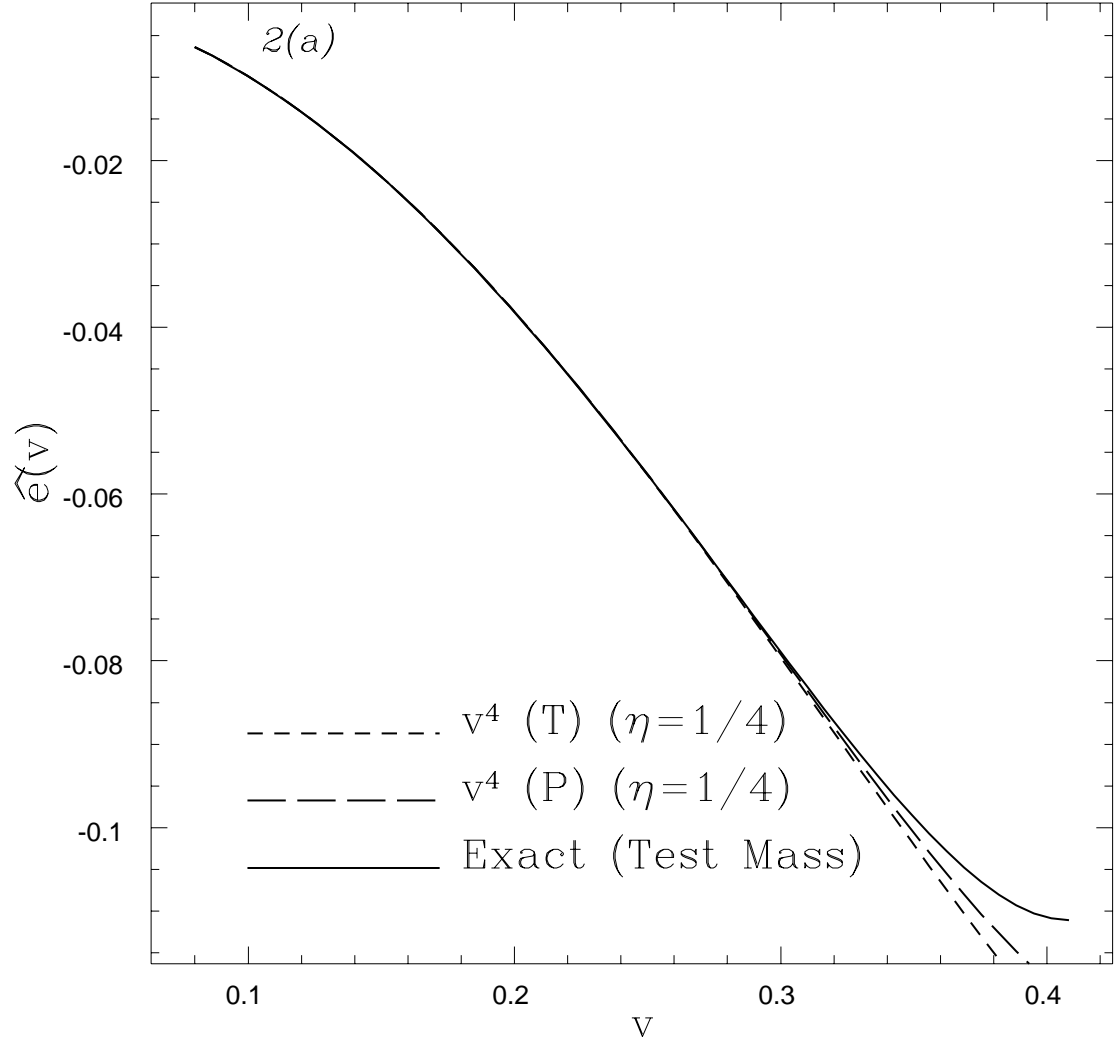} }
\label {fig:energy1}
\end {figure}

Having argued that $e_{P_4} (x)$, Eq. (\ref{eq:n29}), and the corresponding
$E_{P_4} (x)$ defined by Eq. (\ref{eq:Etildecf}), are better estimates of the
finite-mass energy functions than their straightforward post-Newtonian
approximations, Eqs. (\ref{eq:n18}), (\ref{eq:18}), we can use our results
so far
to estimate both the location of the last unstable circular orbit (light
ring) and
that of the last stable circular orbit. The functions $\hat e_{P_4} (v)$,
$\hat E_{P_4} (v)$ are plotted in Fig.~\ref{fig:energy1} together with
$\hat e_{T_4}(v)$ and $\hat E_{T_4}(v)$, both sets for $\eta=1/4$,
and compared with the exact functions $\hat e(v)$ and
$\hat E'(v)$ in the $\eta=0$ (i.e. test mass) case. 
We see that the $\eta=1/4$ $P$- and $T$-approximants are 
smooth deformations of their test-mass limits. 
Note that the variable $x\equiv v^2$ is, in the limit $\eta \rightarrow 0$,
equal to $m/r$ in Schwarzschild coordinates and can be used as a smooth radial
coordinate. If we wished we could also introduce the function $J_{\rm tot} (x)$
giving the $x$-variation of the total angular momentum. It is indeed related to
the total energy $E_{\rm tot} (x)$ by the general identity (for circular orbits)
$d \, E_{\rm tot} = \Omega \, d \, J_{\rm tot}$ where the circular frequency is
given by $m \, \Omega = v^3 = x^{3/2}$. The consideration (even without knowing
its precise analytical form) of the effective potential for general (non
circular)
orbits $E_{\rm tot} = E_{\rm tot} (r,J_{\rm tot})$ in terms of any smooth
radial-type variable $r$ measuring the distance between the two bodies
allows one
to see (by smooth deformation from the $\eta =0$ case) that the minimum of
$E_{\rm
tot} (x)$ (which necessarily coincides with the minimum of $J_{\rm tot} (x)$)
defines the last stable circular orbit. Indeed, it is the confluence of the
one-parameter sequence of {\it minima} of $E_{\rm tot} (r,J_{\rm tot})$
considered
as a function of $r$ for fixed $J_{\rm tot}$ (stable circular orbits) with the
one-parameter sequence of {\it maxima} of $E_{\rm tot} (r,J_{\rm tot})$
(unstable
circular orbits). Note also, from Eq. (\ref{eq:Etildecfprime}), that the last
stable orbit (minimum of $E(x)$) necessarily coincides with the minimum of the
function $e(x)$. As for the last unstable circular orbit it is clearly
defined by
the square-root singularity $\propto (x-x_{\rm pole})^{-1/2}$ of $E(x)$,
corresponding to a simple pole $(x-x_{\rm pole})^{-1}$ in $e(x)$. Applying these
general considerations to our specific $2PN$-Pad\'e proposal (\ref{eq:n29}) one
easily finds that we predict the following ``locations'' (in the invariant $x$
variable) for both the light ring (corresponding to $r=3m$ for a test mass
around
a Schwarzschild black hole),
\begin{equation}
x_{\rm light \, ring}^{P_4} (\eta) = x_{\rm pole}^{P_4} (\eta) = \frac{1}{3} \,
\frac{\left (1+\frac{1}{3} \, \eta \right )}{\left (1-\frac{35}{36} \, \eta
\right) } \, , \label{eq:n30}
\end{equation}
and for the last (circular) stable orbit,
\begin{equation}
x_{\rm lso}^{P_4} (\eta) = \frac{1}{6} \, \frac{\left( 1+\frac{1}{3} \, \eta \right)}{\left (1-\frac{35}{36} \, \eta \right )} \,
\left[ 2 - \frac{\left( 1+ \frac{1}{3} \, \eta \right)}
{\sqrt{1-\frac{9}{16} \, \eta + \frac{1}{36} \, \eta^2}} \right] \, .
\label{eq:n31}
\end{equation}

We recall that $x$ is invariantly defined in terms of the orbital circular
frequency $\Omega = 2\pi \, f^{\rm orb}$ through $x=(m \, \Omega)^{2/3}$,
so that
the gravitational wave frequency (twice the orbital frequency) reads
\begin{equation}
f^{\rm GW} = 2 f^{\rm orb} = \frac{x^{3/2}}{\pi m} = 4397.2 \,
(6x)^{\frac{3}{2}} \,
\frac{m_{\odot}}{m} \, {\rm Hz} \, . \label{eq:n32}
\end{equation}
In the equal mass case $(\eta = 1/4)$ Eqs. (\ref{eq:n30}), (\ref{eq:n31})
yield $3
x_{\rm pole}^{P_4} \left( \frac{1}{4} \right) = 156/109 \simeq 1.4312$, $6
x_{\rm
lso}^{P_4} \left( \frac{1}{4} \right) \simeq 1.1916$, and therefore $f_{\rm
lso}^{\rm GW} = 2 f_{\rm lso}^{\rm orb} = 5719.4 (m_{\odot} / m) \, {\rm
Hz}$. In
particular, for a $(1.4 m_{\odot} , 1.4 m_{\odot})$ neutron star system, we
predict $f_{\rm lso}^{\rm GW} = 2 f_{\rm lso}^{\rm orb} = 2042.6 \, {\rm
Hz}$. Note
that our estimate of the (invariant) location of the last stable orbit is
significantly different from that of Ref. \cite{KWW}, which estimates, for
instance $f_{\rm lso}^{\rm GW} = 2 f_{\rm lso}^{\rm orb} = 1420 \, {\rm
Hz}$ for a
$(1.4 m_{\odot} , 1.4 m_{\odot})$ system. [Actually, we read on the Figures of
Ref. \cite{KWW} a value $(m \, f_{\rm lso}^{\rm orb})^{\rm KWW} \simeq 0.00963$
which corresponds to $6 x_{\rm lso}^{\rm KWW} \simeq 0.925$ and $f_{\rm
lso}^{\rm
orb} \simeq 698 \, {\rm Hz}$ (instead of $710 \, {\rm Hz}$ marked on their
Figures)
for the $(1.4 m_{\odot} , 1.4 m_{\odot})$ case.] Qualitatively our
$\eta$-dependence is different because we find that $x_{\rm lso}^{P_4} (\eta)$
increases with $\eta$ ($6 x_{\rm lso}^{P_4} (\eta) > 1$ and increasing)
while Ref.
\cite{KWW} estimates a $6 x_{\rm lso}^{\rm KWW} (\eta) < 1$, decreasing with
$\eta$. This is an important physical difference as it means, if we are right,
that binary systems of comparable masses can get closer, orbit faster and emit
more gravitational waves before plunging in than estimated in Ref.
\cite{KWW}. As
said above, we think that the ``hybrid'' approximation used in Ref.
\cite{KWW} is
not reliable, notably because of the strong $\eta$-dependence 
(and consequent increase) of the coefficients in their expansion 
(see also the related criticism of
Ref. \cite{WS}). We think that our approach (in which the expansion coefficients
to $e(x)$ are less strongly modified by $\eta$ and where the crucial coefficient
$a_2$ decreases with $\eta$ which means a larger radius of convergence) is more
likely to indicate the correct trend. We have tried in several ways to test the
robustness of our conclusions under the addition of higher post-Newtonian
corrections to Eq. (\ref{eq:18}). We think, however, that such attempts are not
really conclusive because one does not know in advance what is the ``plausible''
range of values of $3PN$ and higher $\eta$-dependent corrections. [We note
in this
respect that the range considered in Ref. \cite{KWW}, $\vert \alpha_i
\vert_{\max}
= \vert \beta_i \vert_{\max} =10$, is clearly too small as it means, for
instance, a {\it fractional} change in the coefficient of $(m/r)^3$ when $\eta$
changes from $0$ to $1/4$ of $\eta \vert \alpha_3 \vert / 16 < 16\%$, while the
{\it known} fractional change in the coefficient of $(m/r)^2$ is already
$\eta \,
29/12 > 60\%$.] In fact, the relative change (ratio $a_k (1/4)/a_k (0)$
when $\eta$
changes from $0$ to $1/4$) of the successive coefficients in any power
series, such
as the $a_k (\eta)$ in Eq. (\ref{eq:23}) is expected to increase (or decrease)
exponentially with the order $k$ due to an $\eta$-dependent shift of the
convergence radius. For instance, in our case if we write the $3PN$
coefficient as
$a_3 (\eta) = -9 (1+\kappa_3 \, \eta)$ to model the $3PN$ $\eta$-dependence
it is
not meaningful to consider {\it a priori} that $\kappa_3$ can take any values in
the range $\pm \kappa_2 \simeq \pm 1$ (where we introduced $a_2 (\eta) = -3
(1+\kappa_2 \, \eta)$ with $\kappa_2 = -35/36$). As the negative value of
$\kappa_2$ has indicated an increase of the radius of convergence with $\eta$
($x_{\rm pole}^{P_4} (\eta) = a_1 (\eta) / a_2 (\eta) = \frac{1}{3} \, (1+
\kappa_1
\, \eta) / (1+\kappa_2 \, \eta)$ with $\kappa_1 = 1/3$) we would rather
expect a value of $\kappa_3$ such that $a_3 / a_2 \sim a_2 / a_1$, i.e.
$1+\kappa_3
\, \eta \sim (1+\kappa_2 \, \eta)^2 / (1+\kappa_1 \, \eta)$ so that
$\kappa_3 \sim
-1.9$. A value of $\kappa_3$ very different from this estimate (i.e. a value of
$a_3 (1/4)$ very different from $-4.8$) would mean that the coefficient
$a_2 \left(
1/4 \right)$ was accidentally smaller than normal (in which case our
estimates (\ref{eq:n30}), (\ref{eq:n31}) would not be reliable). In
conclusion, we
think that, given the presently available information, our estimates are more
internally consistent than previous ones (which include the relevant works
quoted
in Ref. \cite{KWW}), but that, if $a_2 (\eta)$ is only ``accidentally''
decreased
by turning on $\eta$, they might be off the mark. It will be possible to
make more
precise statements on the reliability of Eq. (\ref{eq:n31}) only when the $3PN$
equations of motion of a binary system are derived (or when numerical
calculations
can reliably locate the last stable orbit). Anyway, we shall see that a
knowledge
of the LSO is not so crucial for extracting the inspiral wave form. [We shall
notwithstanding test below the robustness of our overall approach under possible
uncertainties in the locations of $x_{\rm pole} (\eta)$ and $x_{\rm lso}
(\eta)$.]
This is because: (a) Interferometer noise rises quadratically beyond a certain
frequency; consequently  the noise level is pretty high before light
binaries, such
as  NS-NS and NS-BH, reach the LSO;  only in the case of more massive binaries
consisting  of black holes and/or supermassive stars with total mass in
excess of  25 $M_\odot,$ in the case of initial LIGO, and 60 $M_\odot,$ in
the case of  advanced LIGO, will the frequency at the LSO be in a region
where the detector noise is low. In such cases it is important to know the
location of the LSO accurately because it helps in appropriately truncating
the inspiral wave form in search templates  so that it would not produce
anticorrelation with  the coalescence wave form which is itself not known,
as of now, to any accuracy. In the case of lighter mass binaries what is
really needed is that the  approximate energy function should match the
exact one at frequencies where the detector noise is the least. This is
also true for the flux function as we shall see in the next Section.

\section{Flux function} \label {sec:flux}

Contrary to the case of the energy function where we could draw on a lot of
theoretical information, we have less general {\it a priori} information on the
structure of the flux function $F(v)$. The exact gravitational wave
luminosity $F$
is not known analytically. It has, however, been computed numerically with good
accuracy  in the test particle limit \cite{P95} and we shall use this in
our study.
In the test particle limit the flux is also known analytically to a high
order  in perturbation theory; to order $v^{11}$
\cite{TTS97} we have
\begin{eqnarray}
&&F(v ; \eta =0) =\frac{32}{5} \eta^2 v^{10} \nonumber \\
&&\left [\sum_{k=0}^{11}A_k v^k +
(B_6 v^6  + B_8 v^8+ B_9 v^9 +B_{10} v^{10} +B_{11} v^{11}) \ln v \right],
\label
{eq:N35}
\end{eqnarray}
where the various coefficients $A_k$ and $B_k$ can be read off
from \cite{TTS97},
\begin{eqnarray}
A_0    & = &    1,                \nonumber\\
A_1    & = &    0,                \nonumber\\
A_2    & = &   -3.711309523809524,\nonumber\\
A_3    & = &   12.56637061435917, \nonumber\\
A_4    & = &   -4.928461199294533,\nonumber\\
A_5    & = &  -38.29283545469344, \nonumber\\
A_6    & = &  115.7317166756113,  \nonumber\\
A_7    & = & -101.5095959597416,  \nonumber\\
A_8    & = & -117.5043907226773,  \nonumber\\
A_9    & = &  719.1283422334299,  \nonumber\\
A_{10} & = &-1216.906991317042,   \nonumber\\
A_{11} & = &  958.934970119567,   \nonumber\\
B_6    & = &  -16.3047619047619,  \nonumber\\
B_8    & = &   52.74308390022676, \nonumber\\
B_9    & = & -204.8916808741229,  \nonumber\\
B_{10} & = &  116.6398765941094,  \nonumber\\
B_{11} & = &  473.6244781742307.  \label{eq:36}
\end{eqnarray}

By contrast, in the comparable masses case only the first five Taylor
approximants
of $F(v;\eta)$ are known 
\cite{BDIWW,BDI,WW,BIWW,B96}. 
Explicitly, $B_k (\eta) =0$
($k\leq 5$) and
\begin{eqnarray}
A_0 (\eta) & =   & 1, \nonumber \\
A_1 (\eta) & =  & 0, \nonumber \\
A_2 (\eta) & =  & -\frac{1247}{336} - \frac{35}{12} \, \eta, \nonumber \\
A_3 (\eta) & =   & 4\pi, \;\; \nonumber \\
A_4 (\eta) & =  & -\frac{44711}{9072} + \frac{9271}{504} \, \eta +
\frac{65}{18} \, \eta^2, \nonumber \\
A_5 (\eta) & =  & -\left( \frac{8191}{672} + \frac{535}{24} \, \eta \right)
\, \pi. \label{eq:n36}
\end{eqnarray}

There is, however, a bit of general information about the function $F(v)$ which
can be used to motivate the consideration of a transformed flux function, say
$f(v)$, as a better behaved object. Indeed, as pointed out in Ref.
\cite{CFPS}, the
function $F(v;\eta =0)$ has a simple pole at the light ring ($r=3m$, i.e. $x
\equiv v^2 = \frac{1}{3}$). The origin of this pole is simple to understand
physically in a flat spacetime analog. [It is seen from Refs. 
\cite{P93} and
\cite{CFPS} that the curved-spacetime effects (metric coefficients, Green
function) do not play an essential role and that the origin of the pole can be
directly seen in the source terms, Eqs. (2.14) of Ref. 
\cite{P93}.] Let us
consider two (for simplicity identical) mass points, linked by a relativistic
(Nambu-Goto) string, orbiting around each other on a circle (the string tension
$T$ providing the centripetal force opposing centrifugal effects). One can
easily
find the exact solution of this problem and then estimate the linearized
gravitational waves emitted by the system
\cite{ftn12}.
 Let us keep fixed the rest
masses $m_1
= m_2 = m/2$ and the radius of the orbit $R$ and increase the tension $T$
so that
the particles' velocities $v$ tend to the velocity of light. In this limit, one
finds that $RT \sim p = mv / \sqrt{1-v^2 / c^2}$ and that the gravitational wave
amplitude $h \propto RT + p \sim p$. By taking a time derivative and
squaring one
sees that, as $v \rightarrow c$, the gravitational flux $F \sim \Omega^2 \, h^2
\propto p^2$ tends to infinity like $(1-v^2 / c^2)^{-1}$. This shows that the
finding of Refs. \cite{P93,CFPS} is quite general and that, in
particular,
it is very plausible that a binary system of comparable masses will have a
simple
pole in $F(v)$ when the bodies tend to the light ring orbit. We have seen above
that the light ring orbit corresponds to a simple pole $x_{\rm pole}
(\eta)$ in the
new energy function $e(x;\eta)$. Let us define the corresponding (invariant)
``velocity'' $v_{\rm pole} (\eta) \equiv \sqrt{x_{\rm pole} (\eta)}$. This
motivates the introduction of the following ``factored'' flux function
\begin{equation}
f(v;\eta) \equiv \left( 1-\frac{v}{v_{\rm pole} (\eta)} \right) \, F (v ;
\eta) \,
. \label{eq:n33}
\end{equation}
Note
that multiplying by $1-v/v_{\rm pole}$ rather than $1-(v/v_{\rm pole})^2$
has the
advantage of regularizing the structure of the Taylor series of $f(v)$ in
introducing a term linear in $v$ (which is absent in Eq. (\ref{eq:N35})). Two
further tricks will allow us to construct well converging approximants to
$f(v)$.
First, it is clear (if we think of $v$ as having the dimension of a
velocity) that
one should normalize the velocity $v$ entering the logarithms in Eq.
(\ref{eq:N35}) to some relevant velocity scale $v_0$. In absence of further
information the choice $v_0 = v_{\rm lso} (\eta)$ seems justified (the other
basic choice $v_0 = v_{\rm pole}$ is numerically less desirable as $v$ will
never
exceed $v_{\rm lso}$ and we wish to minimize the effect of the logarithmic
terms).
A second idea, to reduce the problem to a series amenable to Pad\'eing, is to
factorize the logarithms by writing the $f$ function in the form
\begin{equation}
f(v;\eta) = \frac{32}{5} \, \eta^2 \, v^{10} \, \left[ 1+
\ln\frac{v}{v_{\rm lso}
(\eta)} \, \left( \sum_k \ell_k \, v^k \right) + \cdots \right] \, \left[ \sum_k
f_k \, v^k \right] \, . \label{eq:n34}
\end{equation}
The ellipsis in Eq. (\ref{eq:n34}) are meant to represent possible higher powers
of $ \ln \frac{v}{v_{\rm lso}}$. [Such terms do not show up at order
$v^{11}$ when
$\eta =0$ and will be also of no concern when considering the $\eta \not= 0$
results at order $v^6$.] The coefficients $f_k$ are functions of $v_{\rm lso}$
in general\cite{free}.

Finally, we define our approximants to the factored flux function $f(v)$ as
\begin{eqnarray}
&&f_{P_n} (v;\eta) \equiv \frac{32}{5} \, \eta^2 \, v^{10} \nonumber \\
&& \left[ 1+\ln\frac{
v}{v_{\rm lso}^{P_n} (\eta)} \, \left( \sum_{k \geq 6} \ell_k \, v^k \right) +
\cdots \right] \, P_{m+\epsilon}^m \left[ \sum_{k=0}^n f_k \, v^k \right] \, ,
\label{eq:n35}
\end{eqnarray}
where $v_{\rm lso}^{P_n} (\eta)$ denotes the LSO velocity ($\equiv \sqrt{x_{\rm
lso}}$) for the $v^n$-accurate Pad\'e approximant of $e(x)$, and where
$P_{m+\epsilon}^m$ denotes as before a diagonal or subdiagonal pad\'e with $n
\equiv 2m+\epsilon$, $\epsilon =0$ or $1$. The corresponding approximant of the
flux $F(v)$ is then defined as
\begin{equation}
F_{P_n} (v;\eta) \equiv (1-\frac{v}{v_{\rm pole}^{P_n} (\eta)})^{-1} \, f_{P_n} (v;\eta)
\, , \label{eq:nn36}
\end{equation}
where $v_{\rm pole}^{P_n} (\eta)$ denotes the pole velocity defined by the
$v^n$-Pad\'e of $e(x)$. For instance, from Eq. (\ref{eq:n30})
\begin{equation}
v_{\rm pole}^{P_4} (\eta) = \frac{1}{\sqrt 3} \, \left( \frac{1+ \frac{1}{3} \,
\eta}{1-\frac{35}{36} \, \eta} \right)^{\frac{1}{2}} \, . \label{eq:n37}
\end{equation}

Let us now see what this definition gives in practice. In terms of the original
expansion coefficients of $F(v)$, $A_k$ and $B_k$ (considered for any
$\eta$) and
of the fiducial velocity $v_0 \equiv v_{\rm lso}$, coefficients appearing in the
definition (\ref{eq:n35}) read
\begin{eqnarray}
\ell_6 & =  & B_6 \, ,\nonumber \\
\ell_7 & = & 0 \, ,\nonumber \\
\ell_8 & = & B_8 - A_2 \, B_6 \, ,\nonumber \\
\ell_9 & =  & B_9 - A_3 \, B_6 \, ,\nonumber \\
\ell_{10} & = & B_{10} - A_2 \, \ell_8 - A_4 \, B_6 \, ,\nonumber \\
\ell_{11} & = & B_{11} - A_2 \, \ell_9 - A_3 \, \ell_8  - A_5
\, B_6 \,. \label{eq:n38}
\end{eqnarray}
We find (remarkably?) that in the test particle limit the $O(v^9)$ logarithmic
term vanishes identically: $\ell_9 (\eta =0) \equiv 0$. The other
coefficients are
numerically ($\eta =0$; $v_{\rm pole}=1/\sqrt{3}$;
$v_0 = v_{\rm lso} = 1/\sqrt{6}$),
$f_0=    1,$ 
$f_1=   -1.7320508075689,$ 
$f_2=   -3.7113095238095,$ 
$f_3=   18.994547272212,$
$f_4=  -26.694053570105,$
$f_5=  -29.756490254383,$
$f_6=  196.66395901720,$
$f_7= -327.26305863109,$
$f_8=   11.063926928123,$
$f_9= 1188.0521512280,$
$f_{10}=-2884.9014287843,$ and
$f_{11}= 2823.3603070298.$
As for the log-factor in Eq. (\ref{eq:n35}) we find that when it is not 
identically $1$ (i.e. when $n \geq 6$)
it is always smaller than about $1.005$ for $v \leq v_{\rm lso} \simeq
0.40825$ and
much closer to $1$ when $v \lsim 0.2$. Although it is unpleasant to have
logarithms
mixing with powers, they do not seem to introduce, in the present case (after normalization to
$v_{\rm lso}$ and factorization), a serious obstacle to constructing
good approximants to $f(v)$.

\begin {figure}[t]
\caption {Newton-normalized gravitational wave luminosity in the
test particle limit: (a) $T$-approximants and
(b) $P$-Approximants.}
\centerline {\epsfxsize 6 true in  \epsfbox {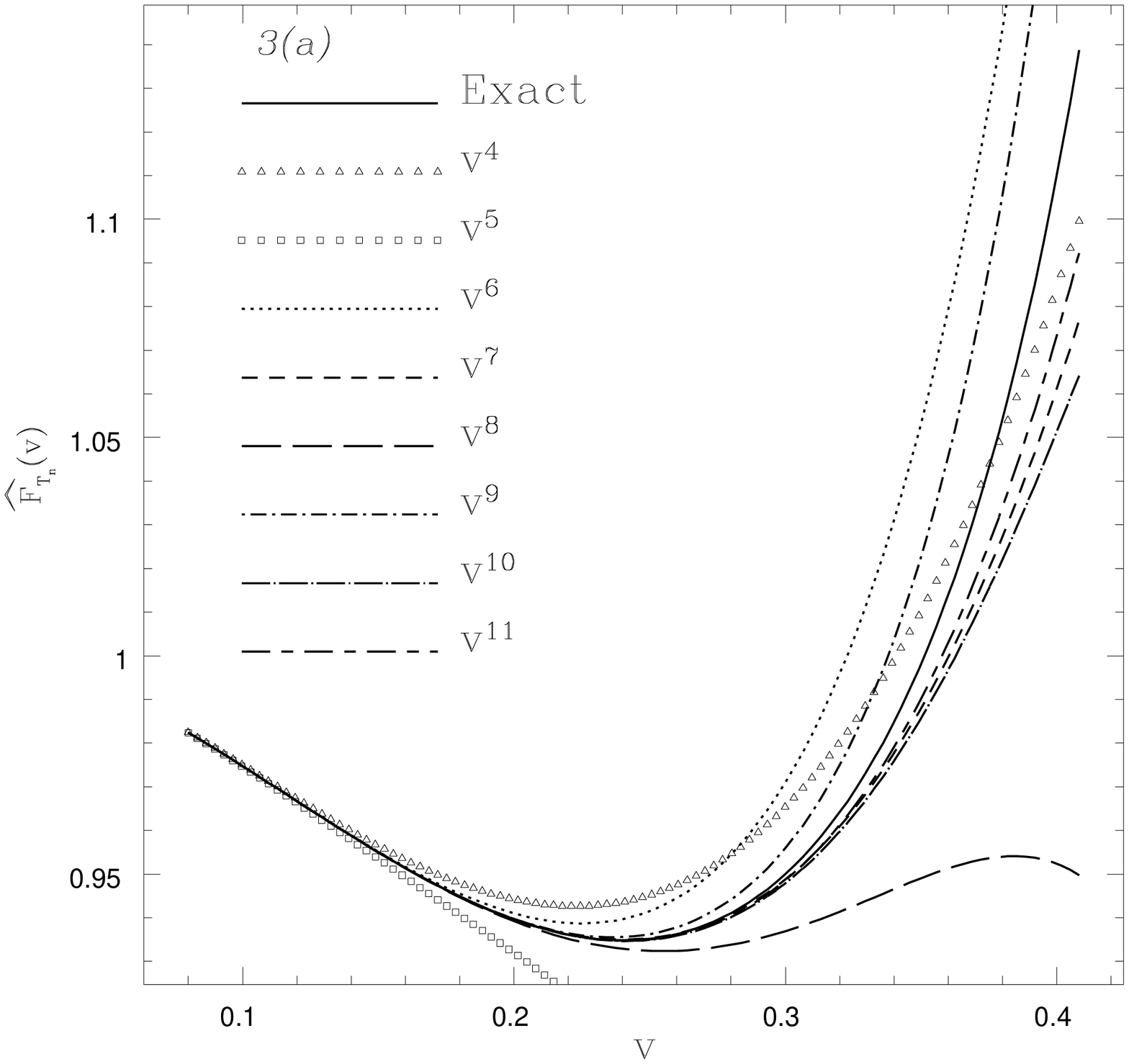} }
\label {fig:flux_tst}
\end {figure}

\begin {figure}[t]
\caption {Newton-normalized factored gravitational wave luminosity 
in the
test particle limit: (a) $T$-approximants and
(b) $P$-Approximants.}
\centerline {\epsfxsize 6 true in  \epsfbox {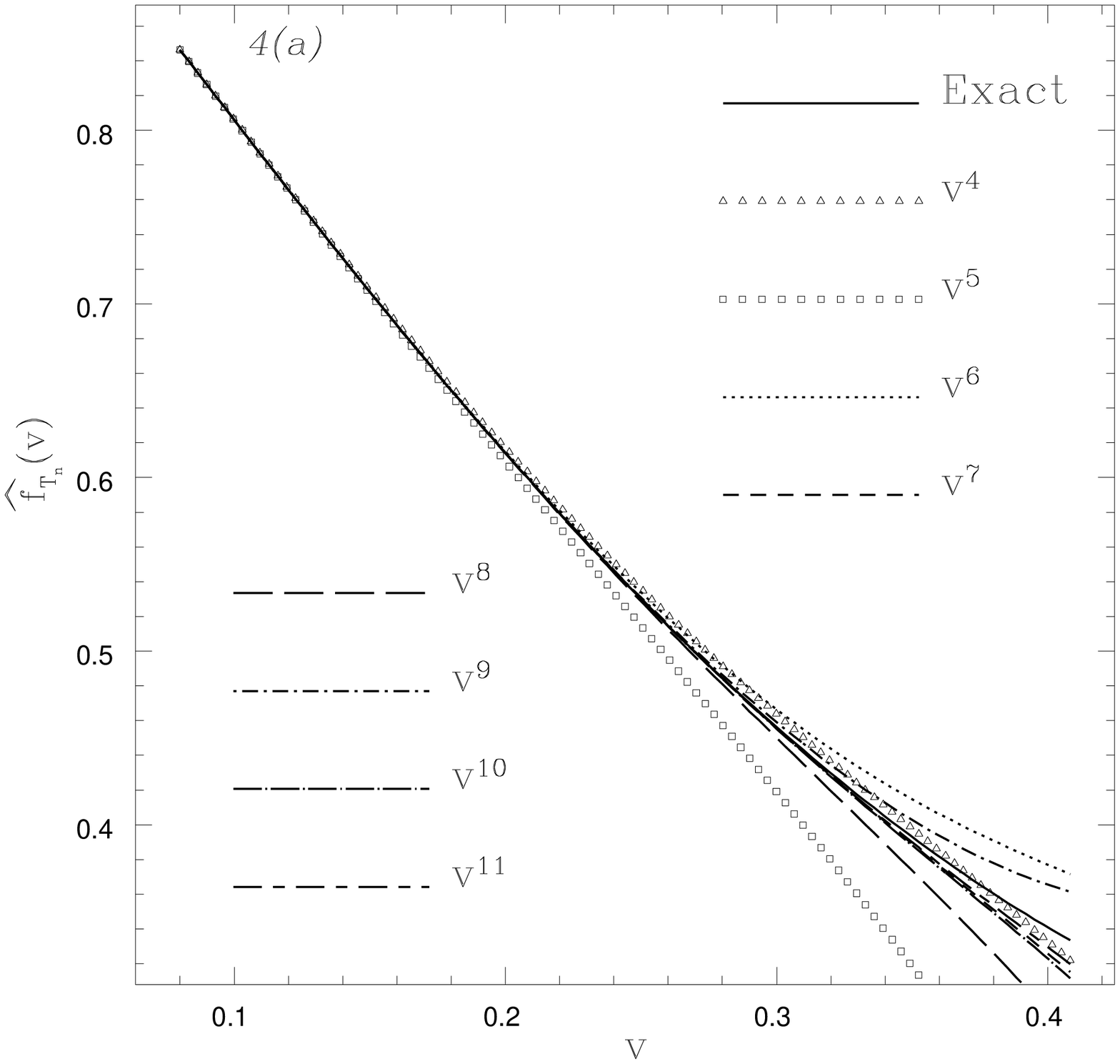} }
\label {fig:flux_tst_factored1}
\end {figure}

Our primary aim in this work is to compare and contrast the convergence
properties
of the standard (``Taylor'') approximants to the phasing formula and its
building
blocks $E(v)$ and $F(v)$ with the new approximants defined above (with their
two-stage construction $E[e_P]$ and $F[f_P]$). Let us first discuss the case of
the flux function which can be studied in detail in the limiting case $\eta
\rightarrow 0$. Indeed, in this case one knows both the ``exact'' (numerical)
flux function \cite{P95}, say $F_X (v)$ and its post-Newtonian expansion up to
order $v^{11}$ \cite{TTS97}. We can then compare directly the approach toward
$F_X (v)$, on the one hand, of the successive standard Taylor approximants
$F_{T_n} (v;\eta =0)$ (obtained by keeping only the $A_k$ and $B_k$ with $k \leq
n$ in Eq. (\ref{eq:N35})) and, on the other hand, of the new approximants
$F_{P_n}
(v;\eta =0)$ defined by Eqs. (\ref{eq:n35}). This comparison of
convergence is illustrated in Fig. \ref {fig:flux_tst}. 
We have plotted there, for convenience, the
``Newton-normalized'' flux functions
\begin{equation}
\widehat{F}_A (v) \equiv \frac{F_A (v)}{F_N (v)} \equiv \frac{5}{32} \,
\eta^{-2} \,
v^{-10} \, F_A (v) \, . \label{eq:n39}
\end{equation}
It is clear that the $P$-approximants converge to the exact values much faster
than the Taylor ones. The monotonicity of the convergence of the
$P$-approximants is also striking. However, the
$P$-approximants of the flux at certain orders
(notably $v^7$ and $v^{10}$) exhibit poles that happen to lie in the
region of integration: $v_{\rm low}<v<v_{\rm lso}.$ Such $P$-approximants
are obviously a bad choice for the construction of templates. 
Nevertheless,
this does not mean that one cannot construct $P$-approximants at that order
at all. Recall that in this study we have only considered diagonal and
subdiagonal Pad\'e approximants of type
$P_m^m$ and $P_{m+\epsilon}^m,$ respectively.
It is perfectly legitimate to employ 
other types of Pad\'e's and in particular the {\it superdiagonal}
Pad\'e of type $P_m^{m+\epsilon}.$ 
For instance, there is a pole in the region of interest in $P^3_4$-approximant
of flux while it turns out that the $P^4_3$-approximant 
(which is the one we have used in this work instead of $P^3_4$)
does not have a pole in the region of interest.
Thus, if one wishes one may trade off a spurious zero,
in the region of interest, in the denominator of the function 
with a zero of the numerator, thereby removing the troublesome pole
(see Appendix~\ref{appendix:pade} for how this may be accomplished 
via some simple properties of the Pad\'e approximants).
For completeness we exhibit in Fig.~\ref{fig:flux_tst_factored1}
the successive $P$-approximants to the factored flux function $f(v;\eta =0)$.

\begin {figure}[t]
\caption {Newton-normalized energy functions
in the comparable mass case. We compare the convergence of the
$T$-approximants and $P$-approximants. Observe 
that the $P$-approximants converge much faster to the fiducial exact
energy than the standard approximants.}
\centerline {\epsfxsize 6 true in  \epsfbox {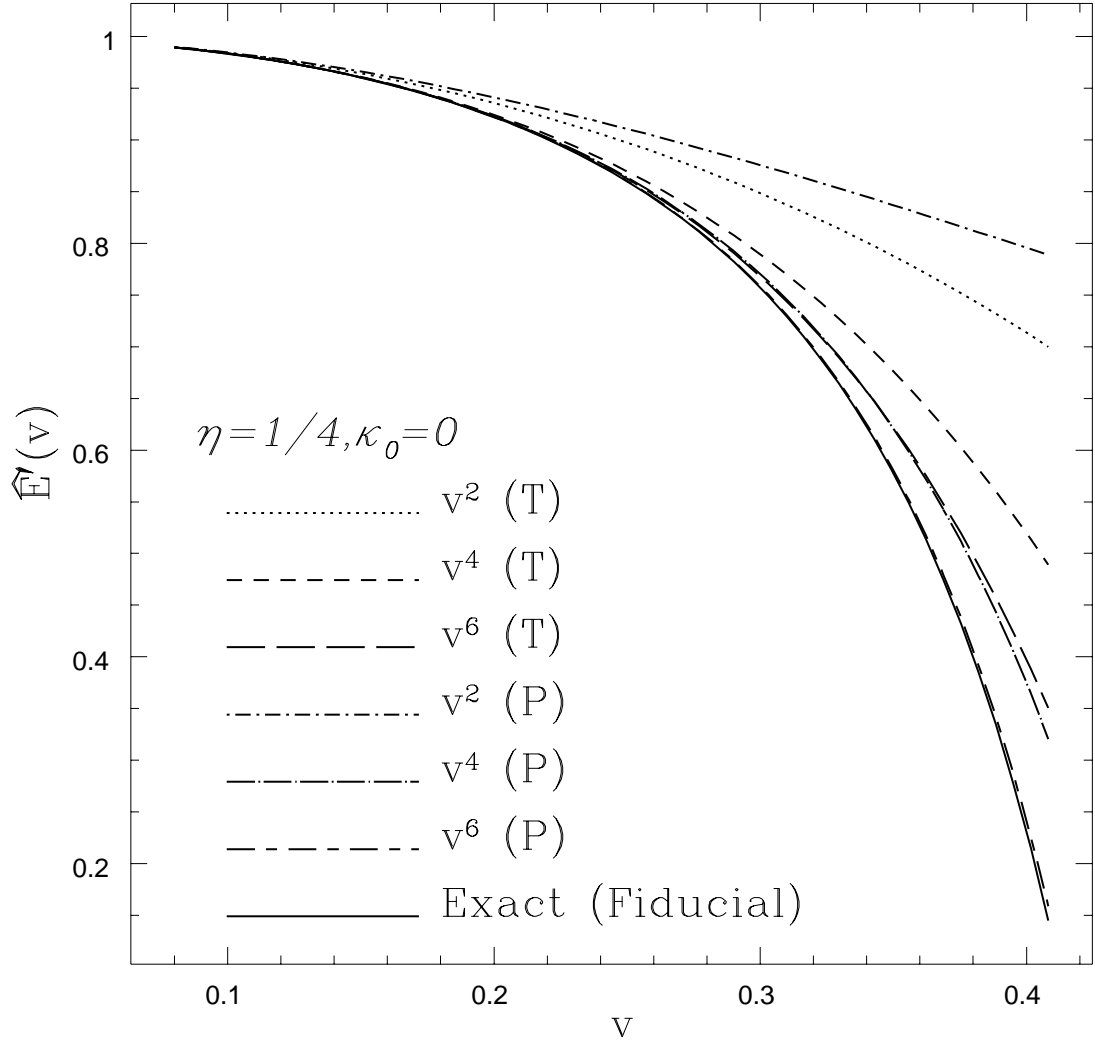} }
\label {fig:energy2}
\end {figure}

\begin {figure}[t]
\caption {Approximants to basic energy functions $e(v)$
are compared with the fiducial exact energy function. Convergence
of $P$-approximants is apparent.}
\centerline {\epsfxsize 6 true in  \epsfbox {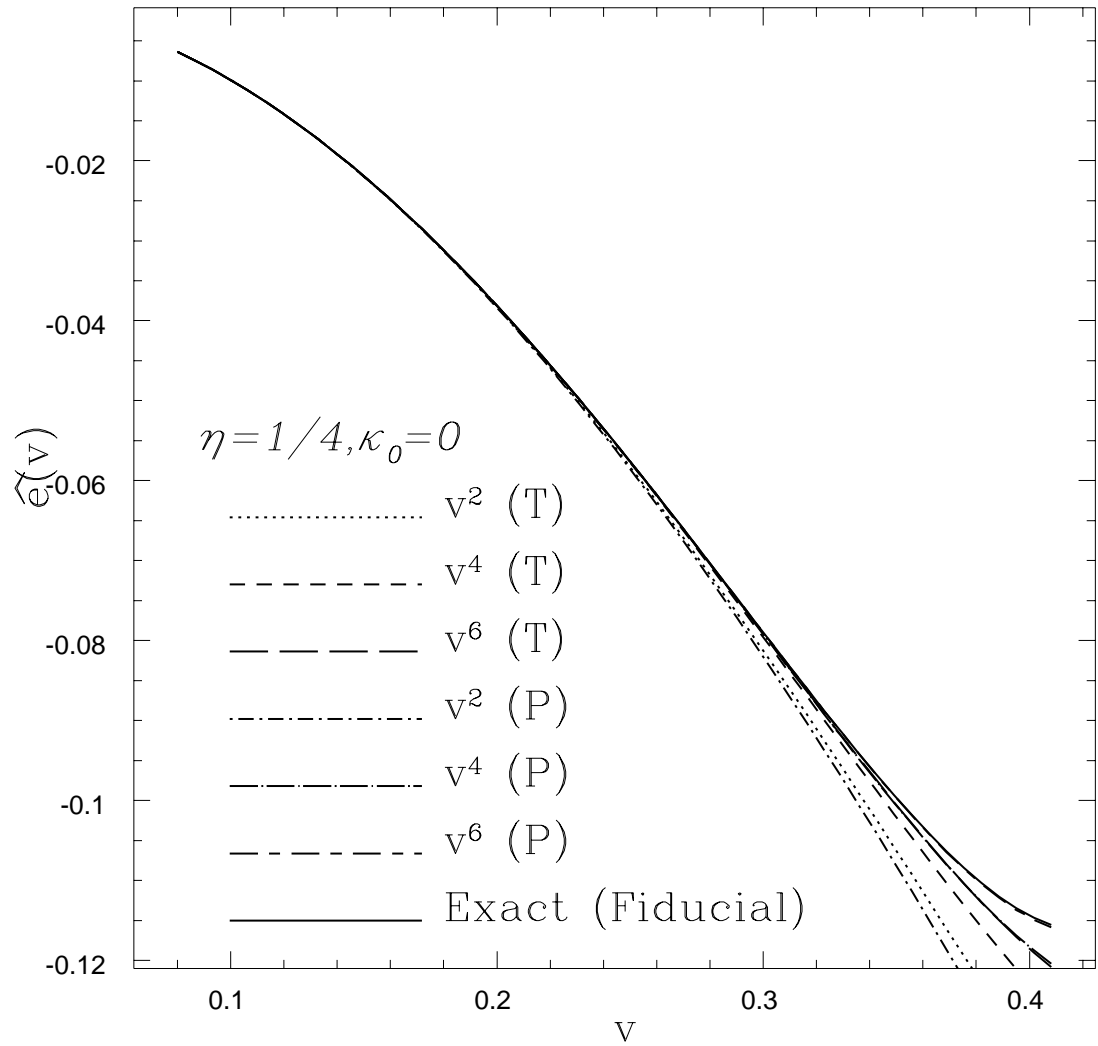} }
\label {fig:basic}
\end {figure}

The other building block of the phasing formula Eqs. (\ref{eq:N10}) are the
approximants to the function $E' (v) = dE(v)/dv$. As we have constructed
$E_{P_n}
(v)$ so that it coincides for $n \geq 4$ with the exact $E_X (v)$ in the case
$\eta = 0$ it would not be fair to compare it to the straightforward $E_{T_n}
(v)$. We need, therefore, to consider the finite
mass case $\eta \not= 0$. However, in this case, we only know few $PN$
approximations and we do not know the exact result. We can formally bypass this
problem and have a first test of the {\it robustness} of our construction by
defining the following fiducial ``exact'' energy function $e_X^{\kappa_0} (x)$:
\begin{equation}
e_X^{\kappa_0} (x; \eta) \equiv -x \left[ 1 - \left( 1+\frac{\eta}{3}
\right) \, x
- \frac{\left( 3-\frac{35}{12} \, \eta \right) \, x^2}{1-3(1-\kappa_0 \,
\eta) \,
x} \right] \, . \label{eq:n40}
\end{equation}
The $2PN$ expansion of $e_X^{\kappa_0} (x)$ coincides by construction with
that of
the ``real'' $e(x;\eta)$. The parameter $\kappa_0$ enters only $3PN$ and higher
order terms. Note that $\kappa_0$ parametrizes an infinite number of $PN$
terms in
a non-perturbative manner because it determines the location of the pole
singularity of $e_X^{\kappa_0}$, namely
\begin{equation}
3 x_{\rm pole}^{\kappa_0} = \frac{1}{1-\kappa_0 \, \eta} \, . \label{eq:n41}
\end{equation}
If we believe our $2PN$ Pad\'e estimate (\ref{eq:n30}) we would expect that
a good
estimate of the ``real'' $\kappa_0$ (when considering $\eta = \frac{1}{4}$)
should
be such that $1-\kappa_0^{P_4} /4 = (1-35 / 144)/(1+1/12)$, i.e.
$\kappa_0^{P_4} =
+47/39 \simeq +1.2051$. To test formally the convergence of the sequence of
$P$-approximants away from the region where we know by construction that it
would
converge very fast we shall consider a value of $\kappa_0$ substantially
different
from the Pad\'e-expected one, for instance simply $\kappa_0 = 0$ which says that
the ``exact'' pole stays when $\eta \not= 0$ at the test mass value $3x=1$
instead
of our result above $3 x_{\rm pole}^{P_4} \left( \frac{1}{4} \right) \simeq
1.4312$. Working again with ``Newton-normalized'' functions, now
\begin{equation}
{\widehat E}'_A (v) \equiv \frac{E'_A (v)}{E'_N (v)} \equiv -\eta^{-1} \,
v^{-1} \,
\frac{d E_A (v)}{dv} \, , \label{eq:n42}
\end{equation} we compare in Fig. \ref {fig:energy2} the convergence of ${\widehat
E}'_{T_n} (v)$
and ${\widehat E}'_{P_n} (v)$ toward the fiducial ``exact'' $e_X^{\kappa_0} (v)$
defined by Eq. (\ref{eq:n40}) for $\eta = \frac{1}{4}$, $\kappa_0 =0$. For
completeness we exhibit also in Fig. \ref{fig:basic} the successive
$P$-approximants to the ``basic'' energy function $e_X^{\kappa_0}$.

The convergence tests performed in this Section have shown at the visual level
that the $P$-approximants behaved better than the $T$-ones. However, the real
convergence criterion we are interested in is that defined by overlaps, to which
we now turn.

\section {Ambiguity Function} \label {sec:ambiguity}

Central to our discussion is the {\it ambiguity function} which is a measure
of the overlap of two wave forms that may differ from each other in not
only their parameter values but also in their shape. For instance, one
of them could be a first post-Newtonian signal corresponding to
non-spinning stars parametrized by masses of the two stars and the other
may be a second post-Newtonian inspiral wave form corresponding to
spinning stars parameterized not only by the masses of the two stars but
also by their spins. Let us therefore consider
two wave forms $h(t;\lambda_k)$ and $g(t;\mu_k)$ where
$\lambda_k,$ $k=1,\ldots,n_\lambda,$ and
$\mu_k,$ $k=1,\ldots,n_\mu,$ are the parameters of the signals and
$n_\lambda$ and $n_\mu$ are the corresponding number of parameters.
The scalar product of these two wave forms is defined in Fourier space by
\begin {equation}
\left < h,g\right >(\tau ;\lambda_k,\mu_k)
\equiv \int_{-\infty}^{\infty} \frac {df e^{2\pi i f \tau}} {S_n(f)}
\tilde h(f; \lambda_k) \tilde g^*(f; \mu_k)
\label {eq:scalarproduct}
\end {equation}
where $\tau$ is the lag of one of the wave forms relative to the other,
$\tilde h (f; \lambda_k)$ and $\tilde g (f; \mu_k)$
denote the Fourier transforms\cite{ftn13}
 of $h(t;\lambda_k)$ and $g(t;\mu_k),$
respectively, $^*$ denotes complex conjugation and $S_n(f)$ is the
two-sided noise power spectral density. The above scalar product is also
the statistics of matched filtering (Wiener filter) which is the strategy
used in detecting inspiralling binary signals. $S_n (f)$ being a (positive)
real,
even function of $f$ the scalar product (\ref{eq:scalarproduct}) defines a real
bilinear form in $h$ and $g$. We introduce also the norm $\Vert h \Vert \equiv
\sqrt{\langle h , h \rangle}$. The {\it ambiguity function} ${\cal A}$ is
defined as
the   value of the normalized scalar product maximized over the lag
parameter $\tau$ :
\begin {equation}
{\cal A}(\lambda_k,\mu_k) = \max_{\tau,\phi} \frac{\langle h,g \rangle
}{\Vert h \Vert \, \Vert g \Vert} \, . \label{eq:ambiguity1}
\end {equation}
where optimization over phases of the signal and the template
is symbolically indicated by $\phi$
(see Appendix~\ref{appendix:phase} for details).
Here $\lambda_k$ can be thought of as the parameters of a signal while
$\mu_k$ those of a template. The signal to noise ratio (SNR) 
for detecting a noise contaminated
version of $h(t)$ with a Wiener filter built from $g(t-\tau)$ reads ${\rm SNR} =
\langle h,g \rangle / \Vert g \Vert$. Its maximum value is ${\rm SNR}_{\max} =
\langle h,h \rangle / \Vert h \Vert = \Vert h \Vert$ when the
time-translated $g$
is perfectly matched to the signal: $g(t-\tau) = h(t)$. Therefore ${\cal A}
(\lambda_k ,\mu_k)$ is the {\it reduction} in SNR obtained using a template
that is
not necessarily matched on to the signal.

The dependences of ${\cal A} (\lambda ,\mu)$ on both $\lambda$ and $\mu$
are important in designing detection strategies. The dependence on the
signal parameters $\lambda$, given some template parameters $\mu$, allow
one to define an optimal way of paving the template parameter space.
The region in the signal parameter space for which a given
template obtains SNRs larger than a certain value \cite {bsssvd91} 
(sometimes called the {\it minimal match} \cite {bjo96}) is 
the {\it span} of that template 
\cite {bjobss97} and the templates should be so chosen that together
they span the entire signal parameter space of interest with
the least overlap of one other's spans. In our case, we are mainly
interested in keeping the signal parameters $\lambda$ fixed, and varying
the template ones $\mu$. In searching for a coalescing binary signal in the
output of a detector one maximizes over a given bank of templates (i.e.
over a dense lattice of $\mu$ values). Thus, the quantity of interest is
the maximum of the ambiguity function over the entire parameter space of
templates.  This maximum, in the case of identical signals, occurs when the
parameters of the template and the signal are equal and is equal to 1.
However, in reality
the template wave forms are not identical to the fully general relativistic
signal and hence the maximum overlap will in general be less than 1
(Schwarz inequality) and would occur not when the parameters are matched
but when they are mismatched:
\begin{equation}
\max_{\mu_k} {\cal A}\le 1.
\end{equation}
If the template wave forms are not `close'
to signal wave forms then it is reasonable to expect that the
maximum occurs when $|\lambda_k-\mu_k|$ is fractionally rather large.
In this case there is not only a substantial reduction in the maximum
SNR that can be achieved by using such a bank of templates but there would
also be a large systematic bias in the measurement of parameters. Using the
terminology of the Introduction such template wave forms would be neither
effectual nor faithful. For {\it detection} purposes we wish to construct {\it
effectual} templates, i.e. templates having large overlap after
maximization over
$\mu$. For {\it parameter estimation} we further need to construct {\it
faithful}
templates which have large overlaps when $\mu \simeq \lambda$. A practical (non
rigorous) criterion for faithfulness is that the ``diagonal'' ambiguity
function
${\cal A} (\lambda ,\lambda)$ be close to 1.

Reduction in the overlap of template wave forms and true signals 
has an effect on the number of detectable events or, equivalently, 
loss in the detection probability of a signal of a given strength. 
For a given signal-to-noise ratio, the distance up to which a detector
can see depends primarily on the amplitude $h_0$ of the wave.
Unavailability of a copy of the true signal means that the
effective strength of the signal reduces from $h_0$ to ${\cal A} h_0$
and hence the span of a detector reduces by the factor ${\cal A.}$
The number of events a detector can detect being proportional to
the cube of the distance, a reduction in the overlap by a factor
${\cal A}$ means a drop in the number of detectable events,
as compared to the case where a knowledge of the true wave form
was available, by a factor ${\cal A}^3.$ For instance, a 10\% (20\%) 
loss in the overlap would mean a 27\% (50\%) loss in the  number of 
events \cite {bjo96}. 
The aim of PN calculations is to make this overlap  as
close to 1 as possible. If we demand that we should be able to detect with PN
templates about 90\% (99\%) of the signals that we would  detect had we 
known the general relativistic signal, then we should have the 
overlap to be no less than about 0.965 (0.997).

As a model for noise we use the expected noise power spectral density
in the initial LIGO interferometer \cite {cf94}:
\begin {equation}
S_n(f) = \frac{S_0}{\alpha+3} \left [ \alpha +
2 \left ( \frac{f}{f_0} \right )^{2} +
  \left ( \frac{f}{f_0} \right )^{-4}  \right ], \ \  f>f_s
\end {equation}
where $S_0,$ $\alpha$ and $f_0$ are constants that characterize
the detector sensitivity, effective bandwidth and the frequency at which
the detector noise is the lowest, respectively. In the case of 
initial LIGO $\alpha=2,$ $f_s=40$~Hz and $f_0=200$~Hz. Due to the 
fact that the noise is essentially infinite below the seismic 
cutoff $f_s$ and since we terminate the template wave forms when 
the velocity reaches that of the last stable orbit the overlap 
integral (\ref {eq:ambiguity1}) reduces to
\begin {equation}
{\cal A}(\lambda_k,\mu_k) = \Vert h \Vert^{-1} \, \Vert g \Vert^{-1} \,
\max_{\tau,\phi}
\left [2 \int_{f_s}^{f_{\rm lso}} \frac {df e^{2\pi i f \tau}} {S_n(f)}
\tilde h(f; \lambda_k) \tilde g^*(f; \mu_k)\right],
\label{eq:ambiguity2}
\end {equation}
where $f_{\rm lso}$ is the gravitational wave frequency 
corresponding to the last stable orbit.  In order to compute 
the maximum overlap we proceed in the following manner. The 
evolution of phase as a function of time is obtained by inverting 
numerically $v$ in terms of $t$ from Eq. (\ref{eq:n10b}) and 
inserting the result in Eq. (\ref{eq:n10c}) and then (\ref{eq:n10a}). 
Though, the iterative procedure in inverting $v$  in terms of 
$t$ is rather computationally intensive, we need to employ it 
since the inaccuracies introduced by the stationary phase
approximation in computing the Fourier transform of the wave form
increase with the  order  of approximation especially in the case
of NS-BH and BH-BH binaries. In Table \ref {stationary.phase}, we give
a measure of the inaccuracies introduced by the stationary phase
approximation at various post-Newtonian orders by computing the
integral in Eq. (\ref {eq:ambiguity2}) with $\tilde h(f)$ 
being the fast Fourier transform
and $\tilde g(f)$ being the stationary phase approximation
(The three cases $A_0,$ $B_0$ and $C_0$ are defined below).

\section {Results and Discussion} \label {sec:results}

Having in hand the ambiguity function to measure the closeness of two wave
forms\cite{ftn14} we can use it to pursue at a quantitative level the analysis of the
convergence of the sequence of approximants defined above.

Let us first consider the wave forms defined in the formal test mass limit where
one keeps the $\eta$ factors in front of $E(v)$ and $F(v)$ but neglects the
$\eta$-dependence in the Taylor coefficients of ${\widehat E}' (v)$ and
$\widehat F
(v)$. Explicitly we mean the wave forms defined by eliminating (numerically) $v$
between \begin{equation}
h^A (v) = C \, v^2 \, \cos 2\Phi_A (v) \, , \label{eq:n43a}
\end{equation}
\begin{equation}
t(v) = t_c - \frac{5}{32} \, \eta^{-1} \, m \int_{v}^{v_{\rm lso}} dv \,
v^{-9} \,
\frac{{\widehat E}'_A (v;\eta =0)}{{\widehat F}_A (v;\eta =0)} \, ,
\label{eq:n43b}
\end{equation}
\begin{equation}
\Phi_A (v) = \Phi_c - \frac{5}{32} \, \eta^{-1} \int_{v}^{v_{\rm lso}} dv \,
v^{-6} \, \frac{{\widehat E}'_A (v;\eta =0)}{{\widehat F}_A (v;\eta =0)} \, ,
\label{eq:n43c}
\end{equation}
in which $v_{\rm lso} = v_{\rm lso} (\eta = 0) = 1/\sqrt 6$. Note that the main
purpose of the overlap computations made for this formal test mass limit is to
{\it compare} quantitatively the convergence of the $P$-approximants to that of
the $T$-ones. One should keep in mind that when studying below in the
formal test
mass limit $(1.4 m_{\odot} , 1.4 m_{\odot})$ or $(10m_{\odot} , 10m_{\odot})$
systems (for which $\eta$ takes its largest value) the absolute values of the
overlaps are not reliable, though one assumes that the lessons learned from the
$P/T$ comparison are. The absolute values of the overlaps for the
$(1.4m_{\odot} ,
10m_{\odot})$ case are probably more reliable, but this is not clear as $\eta
\simeq 0.1077$ is then only a factor 2.32 smaller that $\eta_{\max} = 0.25$.
This being said we wish to compare semi-maximized overlaps that we can
denote for
simplicity as
\begin{eqnarray}
\lefteqn{\langle T_n^0 (m_1 ,m_2), X^0 (m_1 ,m_2)\rangle \equiv} \nonumber \\
& & {\max}_{t_c , \Phi_c^A ,\Phi_c^X} \langle {\widehat h}_n^{T^0} (t_c
,\Phi_c^A , m_1 ,m_2), {\widehat h}^{X^0} (0,\Phi_c^X ,m_1 ,m_2)\rangle\, ,
\label{eq:n44a} \end{eqnarray}
\begin{eqnarray}
\lefteqn{\langle P_n^0 (m_1 ,m_2), X^0 (m_1 ,m_2)\rangle \equiv} \nonumber \\
& & {\max}_{t_c , \Phi_c^A ,\Phi_c^X} \langle {\widehat h}_n^{P^0} (t_c
,\Phi_c^A , m_1 ,m_2), {\widehat h}^{X^0} (0,\Phi_c^X ,m_1 ,m_2)\rangle \, .
\label{eq:n44b} \end{eqnarray}
Here the superscript $0$ on $T$, $P$ or $X$ denotes the above defined
formal $\eta
=0$ limit of Taylor, Pad\'e-type or eXact wave forms, respectively (i.e. $A=T$,
$P$ or $X$ in Eqs. (\ref{eq:n43a})---(\ref{eq:n43c})). 
Here one considers only the same values for
the two dynamical parameters of those signals (i.e. the explicit $m$ and $\eta$
appearing in Eqs. (\ref{eq:n43a})---(\ref{eq:n43c}), 
here expressed in terms of $m_1$ and $m_2$ in
order to psychologically minimize the formal inconsistency of setting $\eta
=0$ in
part of the formula and keeping it elsewhere) and maximize over the kinematical
ones $t_c^A$, $\Phi_c^A$, $t_c^X$, $\Phi_c^X$. To maximize over the reference
times, it is sufficient (as indicated above) to fix $t_c^X =0$ and maximize over
$t_c^A = t_c$ ($\tau$, the time lag). Maximizing over the reference phases
is more
subtle as the overlap depends {\it separately} on $\Phi_c^A$ and $\Phi_c^X$ and
not only on their difference. There is, however, a computationally non-intensive
way to do it which is based on a conceptually simple geometrical formulation 
of the problem (see Appendix~\ref{appendix:phase}).

Note that in Eqs. (\ref{eq:n44a}) the approximate template parameters are not
optimized, but are taken to be equal to that of the exact signal. In other words
we compare the faithfulness of the various approximants together with their
convergence properties. The results are given in Table \ref{faithfulness.tm},
for $n=4-11$ \cite {caution:p43} as well
as for the Newtonian approximants for the purpose of comparison. 
The overlaps quoted are the ${\it minimax}$ overlaps, 
Eq. (\ref{eq:B12}), together with the corresponding {\it best} 
overlaps Eq. (\ref{eq:B11}) in parenthesis below the minimax overlap.
(The $P$-approximant $P_4^3$ corresponding
to $n=7$ has a singularity in the region of interest and hence we
have used the approximant $P_3^4.$ The $P_5^5$-approximant 
too has a pole and we have not computed the overlaps in 
this case though if one desires one can compute other 
$P$-approximants, such as $P_4^6$ or $P^4_6$, at this order.)
We consider three
prototype cases, say case $A_0 [(1.4m_{\odot} ,1.4m_{\odot})]$, case $B_0
[(1.4m_{\odot} , 10m_{\odot})]$, and case $C_0 [(10m_{\odot}, 10m_{\odot})]$. We
added an index zero to recall the fact that $\eta =0$ has been used in
${\widehat
E}'$ and $\widehat F$. [One should keep in mind the warning above that the
numerical
results for case $B_0$ are physically more reliable, while $A_0$ and $C_0$ are
just mathematical ways of testing the convergence.]

We performed another convergence test (still in the formal $\eta \rightarrow 0$
limit) of a different nature. It is known in mathematics that one does not
need to
know in advance the limit of a sequence to test its convergence. One can instead
use Cauchy's criterion which says (roughly) that the sequence converges if,
given
some distance function $d(h,g)$, $d(h_n ,h_m) \rightarrow 0$ as both $n$ and $m$
get large. In our case we have a distance function\cite {measure}
defined by the ambiguity function and we can
compare
the Cauchy convergence of the $T$ and $P$ approximants. Some results are
given in
Table \ref {cauchy.tm} where one exhibits the semi-maximized 
(in the sense of Eqs.  (\ref{eq:n44a})) {\it best} overlaps
$\langle T_n^0 , T_{n+1}^0 \rangle$
versus $\langle P_n^0 , P_{n+1}^0 \rangle$, for $n=4, \ldots, 11,$
and the three prototype cases $A_0$, $B_0$, $C_0$. (As in Table
\ref{faithfulness.tm} where appropriate we have used the
$P_3^4$-approximant instead of $P_4^3.$ Since the $P_5^5$ 
approximant has a pole in the region 
of interest the entries corresponding to $n=10$ are blank and 
the entries corresponding to $n=9$ are the overlaps
$\langle P_9^0 , P_{11}^0 \rangle$.)

The last two Tables show very clearly that the $P$-approximants converge much
better than the $T$-ones and that they provide a much more faithful
representation
of the signal. To measure the {\it effectualness} of our approximants (in the
technical sense defined above) and study the biases they can introduce, we also
performed numerical calculations in which we maximized over all parameters, say
\begin{equation}
\langle\langle T_n^0 ,X^0 \rangle\rangle (m_1 , m_2) \equiv {\max}_{m_1^A ,
m_2^A}
\langle T_n^0 (m_1^A ,m_2^A),X^0 (m_1 ,m_2)\rangle \, , \label{eq:n45a}
\end{equation}
\begin{equation}
\langle\langle P_n^0 ,X^0 \rangle\rangle (m_1 , m_2) \equiv {\max}_{m_1^A ,
m_2^A}
\langle P_n^0 (m_1^A ,m_2^A),X^0 (m_1 ,m_2)\rangle \, , \label{eq:n45b}
\end{equation}
while keeping track of the parameter values $m_1^A , m_2^A$ which, given the
signal parameters $m_1 ,m_2$, maximize the overlaps. The results are
presented in
Table \ref{effectualness.tm}
for the three prototype cases $A_0$, $B_0$, $C_0$
and for the most important values (for the near future) of the order
of approximation: $n=4$, $5$ and $6.$ In this case the overlaps
are the {\it minimax} overlaps.

Our test mass results sum up the general behavior of the different approximants
pretty well. First let us note that {\it even at $O(v^{11})$ $T$-approximants
do not achieve the requisite overlap of 0.965} except in the case of light
binaries. This is consistent with the concern often expressed in the literature
about the need for higher order post-Newtonian wave forms.
In our view the most worrying aspect of the $T$-approximant is not that it
does not obtain a high overlap but that the behavior of the approximant
is oscillatory in nature. For instance, the $O(v^6)$ $T$-wave form achieves an
overlap, with the exact wave form, of about 0.96 which reduces at $O(v^8)$
to as low as 0.71 for system $B_0$ and 0.85 for system $C_0$ (though for system
$A_0$ it maintains a level of 0.965), increases at $O(v^9)$ to about 0.93
for these systems and again drops back at $O(v^{11})$ to 0.85 and 0.90 for
systems $B_0$ and $C_0,$ respectively. One clearly notices that $P$-approximants
do not show such an erratic behavior. Recall that, in the test mass case, 
we are
comparing a {\it known} exact wave form with 
an approximate signal model and hence
the above conclusions are free from any prejudice.
Though the second post-Newtonian $P$-wave form is not a faithful signal model, 
at 5/2 post-Newtonian order the $P$-approximant is a faithful signal model.

Moreover, $P$-approximants show an excellent Cauchy convergence as evident from
Table \ref{cauchy.tm}. Notice that the $T$-approximants have a poor Cauchy
convergence for systems $B_0$ and $C_0.$  This makes them ill-suited as faithful
templates. $T$-approximants are not always effectual signal models either. Sometimes
they do obtain overlaps larger than 96.5\% but at the cost of producing a very large
bias in the estimation of total mass. This is to be 
contrasted with the
$P$-approximants which are effectual at $O(v^4)$ at the level of 99.7\% or better
at the cost of very little bias ($\delta m/m$ always less than 3.5\% and 
less than 1\% in most cases). We have also computed the biases in
the estimation of the parameter $\eta$ and there too we see a similar
trend.

\section{Robustness} \label{sec:robustness}

Up to this point in the paper we have mainly relied on the test mass limit to
assess the quality of our approximants. In this section we shall try to go
beyond
this formal limit to check the {\it robustness} of our proposal under the
turning
on of $\eta$.

\begin {figure}[t]
\caption {Newton-normalized gravitational wave luminosity 
in the comparable mass case. Curves are plotted for three values of
the mass ratio (0,  14/129.96 and 1/4) with thicker curves corresponding
to larger values of the mass ratio. 
(a) $T$-approximants and
(b) $P$-Approximants. For comparison we have also plotted the test particle
flux. Note that the $T$-approximants 
as well as the $P$-approximants, are continuous deformations 
of the test mass limit.}
\centerline {\epsfxsize 6 true in  \epsfbox {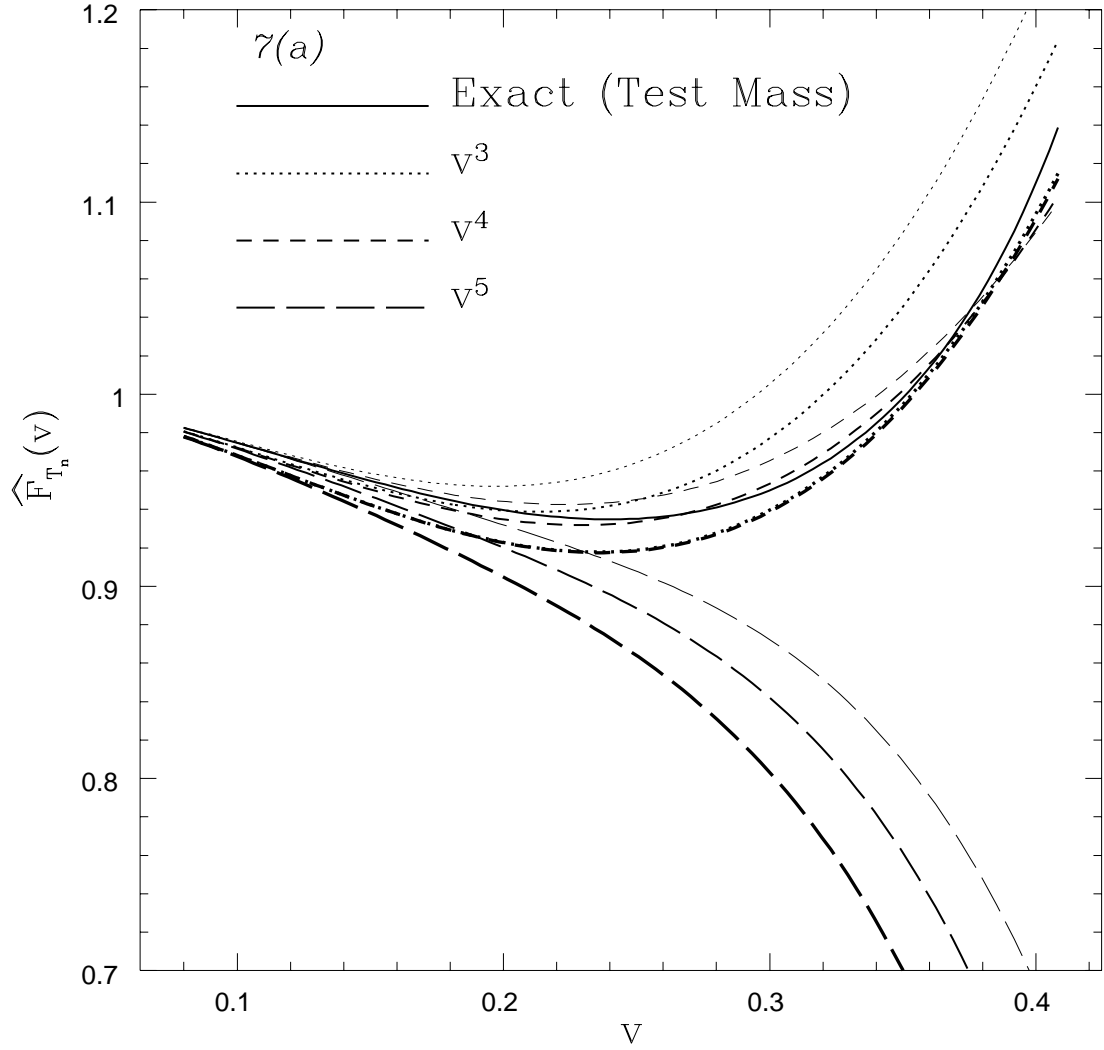} }
\label {fig:flux_tst_factored2}
\end {figure}

We can first use all the existing information about the comparable masses
case and
see whether turning on $\eta$ modifies in any way the trend we saw above. As a
first test (a ``visual'' one) we plot in Fig. \ref{fig:flux_tst_factored2} 
the Newton-normalized flux
functions, ${\widehat F}_{T_n} (v;\eta)$, ${\widehat F}_{P_n} (v;\eta)$ as a
function of $v$, for the maximal value $\eta = \frac{1}{4}$ and for the
cases where
we know them, i.e. $n=2$, $3$, $4$ and $5$. Using the same information we
can also
check the $\eta$-robustness of our Cauchy-convergence criterion. This is done in
Table \ref{cauchy.fm} where we present the semi-maximized {\it best}
overlaps Eq. (\ref{eq:B11})
$\langle P_3^{\eta} (m,\eta) , P_4^{\eta} (m,\eta) \rangle$, 
$\langle P_4^{\eta} (m,\eta) , P_5^{\eta} (m,\eta) \rangle$ and
compare them to their $T$-counterparts for the (real) cases $A,$ $B$ and $C.$

We also made many attempts at testing the robustness of our conclusions when
taking
into account the existence of (unknown) higher-order $\eta$-dependent
corrections.
There is no really conclusive way of achieving such a task but here is our best
attempt: Our starting point is to model an infinite number of (unknown)
higher-order $PN$ corrections by just one (non perturbative) parameter:
$\kappa_0$. As introduced in Eq. (\ref{eq:n40}) above, $\kappa_0$
parametrizes our
ignorance about the true location of the light ring (pole in $e(x)$ and $F(v)$).
Our $2PN$ Pad\'e estimates gave us an $\eta$-corrected value $v_{\rm pole}$, but
we wish to consider here the possibility that maybe the true value 
is quite different from our
estimate. More precisely Eq. (\ref{eq:n40}) parametrizes the pole at $3 x_{\rm
pole} = (1-\kappa_0 \, \eta)^{-1}$, while $3x_{\rm pole}^{P_4} \simeq
1.4312$ for
$\eta = \frac{1}{4}$ corresponding to $\kappa_0 \simeq + 1.2051$. To explore a
very large range of possibilities we shall consider that the true value of
$\kappa_0$ (for $\eta = 0.25$) might range between $\kappa_0 =-1$ (meaning
$3x_{\rm pole} =0.8$) and $\kappa_0 = +2$ (meaning $3x_{\rm pole} =2.0$).
In Table~\ref{lso.fm} we compare the location of the the last stable orbit
$x_{\rm lso}\equiv v^2{\rm lso}$ predicted by the $T$ and $P$-approximants 
to the energy function relative to the exact location $x^X_{\rm lso}.$ We see
that $P$-approximants capture the location much better than the $T$-approximants.

Having chosen the range of $\kappa_0$ we shall consider, and adopting the
definition (\ref{eq:n40}) for the corresponding fiducial ``exact'' $e$-function,
it remains to define a corresponding fiducial ``exact'' $f$-function, having the
property that the corresponding $F$-function coincides, up to $O (v^6)$
terms, with
the known $T_5$ expansion of $F$. To this effect the simplest proposal is to
define first the $T_{11}$ (Taylor to $v^{11}$) expansion of $f_X^{\kappa_0}
(v)$ by
\begin{eqnarray}
T_{11} [f_X^{\kappa_0}] &\equiv& \left( 1-\sqrt{3(1-\kappa_0 \, \eta)} \, v
\right)
\nonumber \\
{ \ } &{ \ }& \left[ \sum_{k=0}^{5} A_k (\eta) \, v^k + \sum_{k=6}^{11}
(A_k (0) \,
v^k + B_k (0) \, v^k \, {\rm ln} \, v) \right] \, , \label{eq:n46}
\end{eqnarray}
where $A_k (\eta)$, $k \leq 5$, are given by Eq. (\ref{eq:n36}), and the others
$(\eta = 0)$ by Eq. (\ref{eq:36}). Then we define the corresponding fiducial
``exact'' $f$-function by:
\begin{equation}
f_X^{\kappa_0} (v;\eta) \equiv f_{P_{11}} (v;\eta) \equiv \, \hbox{the R.H.S. of
Eq. (\ref{eq:n35}).} \label{eq:n47}
\end{equation}
Having defined some fiducial ``exact'' $e$ and $f$ functions we have
correspondingly defined some ``exact'' wave form $h_X^{\kappa_0}$ and, using the
definitions above, both $T$-type and $P$-type approximants of this wave form. We
are interested in knowing whether the $P$-approximants behave better than the
$T$-ones even in presence of higher-order effects significantly different
from the
behavior expected from the 2PN Pad\'e results. 
The results of this exercise are
presented in Table \ref {faithfulness.fm} where one has computed 
the semi-optimized {\it minimax} overlaps
$\langle P_n^{\eta} (m,\eta), X_{\kappa_0}^{\eta} (m,\eta)\rangle$ and 
$\langle T_n^{\eta} (m,\eta),X_{\kappa_0}^{\eta} (m,\eta) \rangle$ 
for the cases $A$, $B$, $C$, for
$\kappa_0 = -1,$ $1.2051$ and $2$ and for $n=4,$ 5, 6 and 7.
In order to test the effectualness of the approximants, in 
Table \ref{effectualness.fm} we quote the fully-optimized but
{\it minimax} overlaps
$\langle \langle P_n^{\eta} (m,\eta),X_{\kappa_0}^{\eta} (m,\eta)\rangle \rangle$ and 
$\langle \langle T_n^{\eta} (m,\eta),X_{\kappa_0}^{\eta} (m,\eta)\rangle \rangle$ 
again for the cases $A$, $B$, $C$, for
$\kappa_0 = -1,$ $1.2051$ and $2$ and for $n=4,$ 5, 6 and 7.

From Table \ref {faithfulness.fm} we clearly see that 
$T$-approximants fail to be faithful signal models even at the
third post-Newtonian order. The second post-Newtonain wave form
of this family would clearly fail to capture even 20\% of all
potential NS-NS events that would be detectable with the aid of
a family of templates constructed out of $P$-approximants.
Even when parameter values are extreme ($\kappa_0=-1,$ and very high
masses) the presently available 5/2 post-Newtonian energy and flux
functions are sufficient to construct a faithful $P$-approximant.

We observe that except when the parameter values are extreme (very low value
of $\kappa_0$ and high masses) $O(v^5)$  $P$-approximants are indeed 
good effectual signal models. In fact in all cases, but one, they obtain
an overlap in excess of 99\%. Bias in the estimation of 
the total mass is at worst 7.6\% and in many cases it is below 2\%. On the
contrary standard second post-Newtonian approximants are not effectual
in many cases; when they are effectual they often produce a relatively large
bias. For instance,  for system $B,$ when $\kappa_0=47/39,$ second
post-Newtonain $T$-approximant acquires an overlap of 0.98 compared to 1.00
acquired by the $P$-approximant of the same order. However, the bias is 97\%
in the former case as compared to a tiny 1.1\% in the latter case. Similarly,
for $\kappa_0=2,$ the 2.5 post-Newtonian $T$-approximant achieves an
overlap of 0.988 at a bias of 75\% while the $P$-wave form 
achieves 0.996 overlap with practically no bias at all.
The biases in the estimation of the $\eta$-parameter (not shown)
are also pretty small when $P$-approximants are used as compared
to $T$-approximants.

A word of caution is in order for those who desire to use standard post-Newtonian
templates: A careful examination of the above Tables reveals that the 
2.5 post-Newtonian $T$-approximant systematically obtains poorer overlaps 
and larger biases. This is of course related to the fact that the 5/2
post-Newtonian flux is very badly behaved (cf. Fig.\ref {fig:flux_tst}).
Hence one must never employ 2.5 post-Newtonian $T$-approximant for searches.
However, $P$-approximants do not suffer from this predicament. Indeed at
$O(v^5)$ $P$-wave form is an excellent effectual signal models. For all
systems and parameters this model obtains an overlap of better than 99.5\%
at a bias less than 1.5\%.

\section{Conclusions} \label {sec:conclusions}

In this work we have studied the convergence properties of various 
post-Newtonian templates to detect gravitational waves emitted by 
inspiralling compact binaries consisting of neutron stars and/or
black holes. We have shown that the standard post-Newtonian
filters, referred to as the $T$-approximants that are
based on Taylor series, considered in the
literature define a badly convergent sequence of approximants. 
Even at order $v^{11}$ the $T$-approximants
only provide overlaps $\sim 0.86$ with the exact signal
in the case of binaries consisting of 1.4-10 $M_\odot$ systems. 
Worse, the convergence of the sequence
of $T$-approximants is oscillatory rather than monotonous. 
Our results on $T$-approximants confirm 
previous, less convincing arguments in the literature, which were 
either based on rough quantitative estimates, or on 
numerical calculations based on the stationary phase approximation 
for Fourier transforms --- an approximation that we have 
shown not to be sufficiently accurate for this purpose
(see Table \ref {stationary.phase}).

\begin {figure}[t]
\caption {Histogram of fraction of events accessible using
$T$- and $P$-approximants relative to the case when the
true signal is known.}
\centerline {\epsfxsize 6 true in  \epsfbox {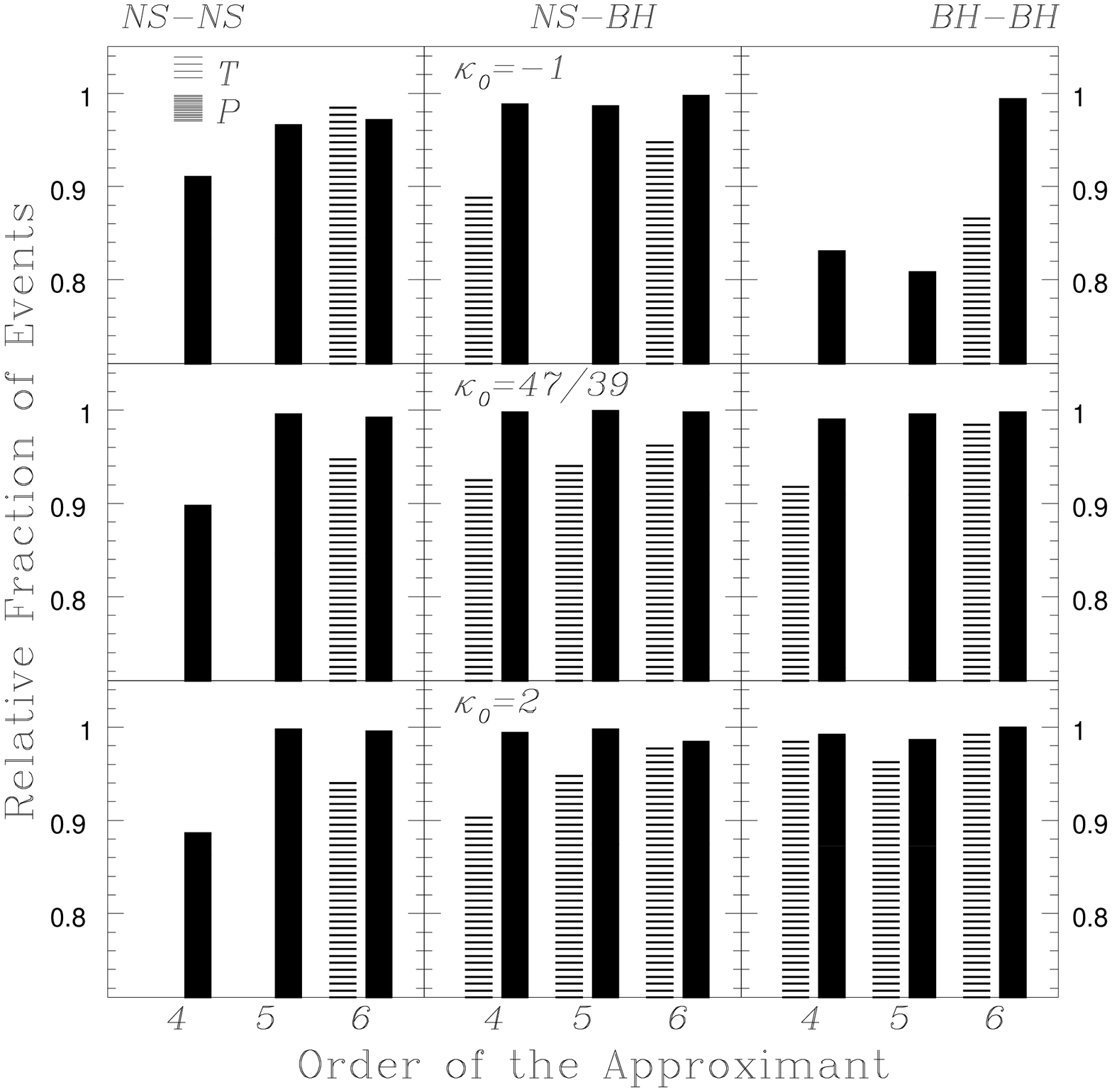} }
\label {fig:eventrate}
\end {figure}

We have defined a new sequence of approximants, referred
to as the $P$-approximants, based on two ingredients:
(i) the introduction, on theoretical ground, of two new energy-type and 
flux-type functions $e(v)$ and $f(v)$, instead of the conventionally
used $E(v)$ and $F(v)$ and (ii) the systematic use of Pad\'e 
approximation for constructing successive approximants
of $e(v)$ and $f(v).$ The new sequence of $P$-approximants has 
been shown to exhibit a systematically better convergence behavior 
than the $T$-approximants.
The overlaps they achieve at a fixed post-Newtonian order 
are usually much higher, and the 
convergence is essentially monotonous instead of oscillatory (as
pictorially described in Fig~\ref{fig:conceptual}
and mathematically measured by the overlaps quoted
in Tables \ref {faithfulness.tm},
\ref {effectualness.tm}, \ref {faithfulness.fm}, 
and \ref {effectualness.fm}). From our 
extensive study of the formal ``test-mass limit" 
$\eta \equiv m_1m_2/(m_1+m_2)^2 \Rightarrow 0,$
i.e. keeping overall $\eta$-factors but neglecting $\eta$
in the coefficients of the post-Newtonian expansions, 
it appears that the presently known $O(v/c)^5$-accurate 
post-Newtonian results allow one to construct approximants having 
overlaps larger than 96.5\% (overlaps corresponding 
to $\kappa_0=47/39,$~2 in Table \ref{effectualness.fm} and
all, but one, overlaps in Table \ref{effectualness.fm})
with the exact signals. 
Such overlaps are enough to guarantee that no 
more than 10\% of signals may remain undetected.
By contrast $(v/c)^5$-accurate $T$-approximants only give 
overlaps of 50\%, and sometimes even as low as 30\%,
corresponding to a loss of 87.5\% and 97\% events, respectively.
Our results are summarized in Fig.~\ref{fig:eventrate}
where we have plotted the fraction of events which
the templates constructed out of $T$ and $P$-approximants
would detect relative to the total number of events that
would have been detectable if we have had access to the true signal.
We clearly notice the superiority of the $P$-approximants. 
Moreover, our computations indicate that the new templates 
entail only acceptably small biases in the estimation of 
signal parameters (see Tables \ref{effectualness.tm} 
and \ref{effectualness.fm}).  In the terminology introduced 
in the text, $P$-approximants are both more
effectual (higher fully maximized overlaps), 
and more faithful (smaller biases) than the usual $T$-approximants.
The above conclusions are primarily based on the study of the 
formal test-mass limit and assumes that turning on $\eta$
brings only a smooth deformation of what happens at $\eta\Rightarrow 0.$ 
We have also studied the effect of turning on $\eta$
($\eta \ne 0$) in the coefficients of the post-Newtonian expansions. 
From all our checks it seems that the $\eta$-dependence 
is indeed smooth and should not alter the fact the 
$P$-approximants have a better convergence
than the $T$-ones. Our construction predicts that the 
last stable circular orbit is closer 
(i.e. greater orbital period) when $\eta \ne 0$
(see Eq. (\ref{eq:n31}).  This is good news because it 
improves the efficiency of $P$-approximants to be used 
as filters for detectors having a fixed frequency band. However, 
we have no independent confirmation of this (favorable) dependence
on $\eta.$ We have tested the robustness of our conclusions against
possible very drastic changes brought by (still unknown) 
$\eta$-dependent terms in the higher post-Newtonian coefficients. 
In the case where these extreme
changes go in the opposite direction of what is suggested by
presently known results (i.e. in the case $\kappa_0 = -1$), we find that
the overlaps are worsened compared to our best estimate range 
($\kappa_0 = 47/39$). This shows that it is important to extend 
the presently available $O(v^5)$ post-Newtonian results  
to the third post-Newtonian level (notably for the equations of motion)
\cite{3pn}.
This will allow one to check whether the $\eta$-dependence of the  
2.5 post-Newtonian results that we use is typical of the 
higher terms (as our method assumes)
or exhibit some abnormal behavior for some 
unforeseeable reason.
When third post-Newtonian results are available it is still 
advisable to use the $P$-approximants: they have consistently 
higher overlaps and lower biases (cf. see Table 
\ref {effectualness.fm}).

In this study we have only considered the noise power spectral density
corresponding to initial LIGO interferometers. Naturally, one must study
other cases as well.  Based on the current study we can be confident 
that in all cases the $P$-approximant wave forms will fare much better 
compared to the standard post-Newtonian ones. However, 
their performance in absolute terms needs to be re-assessed 
since other interferometers, such as GEO600, VIRGO, and enhanced LIGO,
happen to have effective bandwidths and the frequency of maximum 
sensitivity somewhat different from initial LIGO. In addition, one
must also address the performance of $P$-approximate wave forms with
regard to parameter estimation.

\vglue 1cm

\section*{Acknowledgments}
It is a pleasure to thank Eric Poisson for providing the numerical test mass
flux.  BRI thanks the Institut des Hautes Etudes Scientifiques
, University of Wales Cardiff and the Albert Einstein Institute, Potsdam,
while BSS thanks the Raman Research Institute and Institute des Hautes
Etudes Scientifique for hospitality during
different phases of this work. This work was supported in part by NSF grant
PHY-9424337. BSS thanks Kip Thorne and his group for useful conversations.

\appendix

\section{Pad\'e Approximants} \label {appendix:pade}

\bigskip

A Pad\'{e} approximant to the truncated Taylor series expansion
of a function is a
rational polynomial with the same number of coefficients as the
latter. The coefficients of the Pad\'{e} approximant are uniquely
determined by reexpanding the Pad\'{e} approximant to the same order
as the truncated Taylor series and demanding that the two agree.
In our study we use a continued fraction form
of the (near diagonal) Pad\'{e} approximant instead of the usual rational
polynomial.

Let $S_n (v) = a_0 + a_1 \, v + \cdots + a_n \, v^n$ be a truncated Taylor
series.
A Pad\'e approximant of the function whose Taylor approximant to order $v^n$ is
$S_n$ is defined by two integers $m,k$ such that $m+k=n$. If $T_n [\cdots]$
denotes the operation of expanding a function in Taylor series and truncating it
to accuracy $v^n$ (included), the $P_k^m$ Pad\'e approximant of $S_n$ is
defined by
\begin{equation}
P_k^m (v) = \frac{N_m (v)}{D_k (v)} \ ; \ T_n [P_m^k (v)] \equiv S_n (v) \, ,
\label{eq:A1}
\end{equation}
where $N_m$ and $D_k$ are {\it polynomials} in $v$ of order $m$ and $k$
respectively. If one assumes that $D_k (v)$ is normalized so that $D_k (0) =1$;
i.e. $D_k (v) = 1+q_1 \, v + \cdots$, one shows that Pad\'e approximants are
uniquely defined by (\ref{eq:A1}). Note that, trivially, $P_0^n [S_n]
\equiv S_n$
which indicates that Pad\'e approximants are really useful when $k\not= 0$.
Actually, it seems that in many cases the most useful Pad\'e approximants
are the
ones near the ``diagonal'', $m=k$, i.e. $P_m^m$ if $n=2m$ is even, and
$P_m^{m+1}$
or $P_{m+1}^m$ if $n=2m+1$ is odd. In this work we shall use,
except when specified otherwise, the diagonal ($P_m^m$)
and the ``subdiagonal'' ($P^m_{m+1}$) approximants. For instance,
the $P^3_4$-approximant of the flux function has a pole and therefore
we use instead the $P^4_3$-approximant.
The diagonal $(P_m^m)$ or subdiagonal
$(P_{m+1}^m)$ Pad\'e approximants can be conveniently written in a continued
fraction form (see e.g. \cite{BO}). For example, given
\begin{equation}
S_2 (v) = a_0 + a_1 \, v + a_2 \, v^2 \, , \label{eq:A2}
\end{equation}
one looks for
\begin{equation}
P_1^1 (v) = \frac{c_0}{1+ \frac{c_1 v}{1+c_2 v}} = c_0 \, \frac{1+c_2 v}{1+(c_1
+c_2) v} \, , \label{eq:A3}
\end{equation}
and given
\begin{equation}
S_3 (v) = a_0 + a_1 \, v + a_2 \, v^2 + a_3 \, v^3 \, , \label{eq:A4}
\end{equation}
one looks for
\begin{equation}
P_2^1 (v) = \frac{c_0}{1+ \frac{c_1 v}{1+ \frac{c_2 v}{1+c_3 v}}} = c_0 \,
\frac{1+(c_2 + c_3) v}{1+(c_1 +c_2 +c_3 ) v + c_1 c_3 v^2} \, . \label{eq:A5}
\end{equation}
The main advantage of using the continued fraction representations is that the
lower order coefficients $c_k$ {\it remain unchanged} as we increase the
order of
the polynomial being approximated. This is not true for the coefficients of the
polynomials $N_m$, $D_k$ in Eq. (\ref{eq:A1}). [This is easily seen by comparing
Eqs. (\ref{eq:A3}) and (\ref{eq:A5}).] The $c_k$'s are algorithmically
obtainable
in terms of the coefficients $a_{\ell}$ in $S_n$ with $\ell \leq k$. For
instance,
\begin{eqnarray}
c_0 & = & a_0 \,, \nonumber \\
c_1 & = & -\frac{a_1}{c_0}, \nonumber \\
c_2 & = & -\frac{c_0 c_1^2 - a_2}{c_0c_1}, \nonumber \\
c_3 & = & -\frac{c_0 c_1 (c_2 + c_1)^2 + a_3}{c_0c_1 c_2}, \nonumber \\
c_4 & = & -\frac{c_0 c_1(  c_2 +  c_1 )^3  +c_0 c_1c_2c_3(c_3+2c_2+2c_1)-a_4 }
         {c_0c_1 c_2c_3}, \nonumber \\
c_5 & = & -\frac{c_0 c_1(( c_2+ c_1)^2 + c_2 c_3)^2+ c_0 c_1 c_2 c_3
         (c_4 +c_3 +c_2 +c_1)^2  +a_5} {c_0 c_1 c_2 c_3c_4}\,, \nonumber \\
c_6 & = & -\frac{1}{c_0c_1c_2c_3c_4c_5}\{c_0c_1(c_1+c_2)^5+ \nonumber\\
        &&c_0c_1c_2c_3[ (c_1+c_2+c_3)^3+3(c_1+c_2)^3+3c_2c_3(c_1+c_2)+c_3(c_2-c_1)]+\nonumber\\
        &&c_0c_1c_2c_3c_4[(c_1+c_2+c_3+c_4)^2+2(c_1+c_2+c_3)^2+c_3(c_4-2c_1)]+\nonumber\\
        &&c_0c_1c_2c_3c_4c_5[c_5+2(c_4+c_3+c_2+c_1)]-a_6\}
\end{eqnarray}
The expressions for the later coefficients are long and not listed here.
Explicitly in terms of the $a$'s some of the above coefficients are ,
\begin{eqnarray}
c_0 & =  & a_0 \, ,\nonumber \\
c_1 & = & -\frac{a_1}{a_0} \, ,\nonumber \\
c_2 & = & -\frac{a_2}{a_1} + \frac{a_1}{a_0} \, , \nonumber \\
c_3 & =  & \frac{a_0 (a_1 a_3 - a_2^2)}{a_1 (a_1^2 - a_2 a_0)} \, .
\label{eq:A6}
\end{eqnarray}

A few other properties of the Pad\'e approximants are useful to notice, such as
\begin{equation}
P_k^m [T_n [f]] = (P_m^k [T_n [f^{-1}]])^{-1} \, , \label{eq:A7}
\end{equation}
\begin{equation}
P_m^{m+\delta} [S_n] = a_0 + a_1 \, v \, P_{m+\delta}^m [\tilde{S}_{n-1}] \, ,
\label{eq:A8}
\end{equation}
where $\delta = 0$ or $1$ and where $\tilde{S}_{n-1}$ is defined by ``factoring
after $v$'': $S_n = a_0 + a_1 \, v \, \tilde{S}_{n-1}$. Eq. (\ref{eq:A8}) shows
how to obtain the superdiagonal Pad\'e of $S_n$ from the continued fraction
approximants $(\equiv P_{m+\delta}^m)$ of $\tilde{S}_{n-1}$. This can be
iterated
to $P_n^{n+2}$ etc$\ldots$ [An alternative way would be, from Eq. (\ref{eq:A7}),
to work with the inverse of the series $S_n$.]

\vglue 1cm

\section { Optimizing over the phases} \label {appendix:phase}

\bigskip

The ``eXact'' (label $X$) and approximate (label $A$) template wave forms
have the
form
\begin{eqnarray}
h^X (t;t_c^X , \phi_c^X ,C^X) & =  & C^X \, a (t-t_c^X) \cos [\phi_c^X + \phi^X
(t-t_c^X)] \, ,\nonumber \\
h^A (t;t_c^A , \phi_c^A ,C^A) & =  & C^A \, a (t-t_c^A) \cos [\phi_c^A + \phi^A
(t-t_c^A)] \, , \label{eq:B1}
\end{eqnarray}
where we denoted $\phi_c^X \equiv 2\Phi_c^X$, $\phi^X (t) \equiv 2\Phi^X (t)$,
etc. The normalized overlap between $h^X$ and $h^A$ depends on the time
difference
$t_c^A - t_c^X$, and (separately) on the two phases $\phi_c^X$, $\phi_c^A$.
Here,
we show how, for any given time lag $\tau = t_c^A - t_c^X$ (i.e. after having
fixed, for instance, $t_c^X = 0$, $t_c^A = \tau$) one can maximize the overlap
over the two phases $\phi_c^A$, $\phi_c^X$.

To solve this maximization problem \cite {bfs} it is useful to think in 
geometrical terms:
each wave form $h(t)$ is seen as a ``vector'' $h$ in an infinite-dimensional
vector space ${\cal W}$ endowed with the Euclidean metric defined by the
(Wiener)
scalar product. For any given $t_c^X$, $t_c^A$, one sees by expanding the
cosines
by the usual addition formula that the two-parameter family of ``vectors'' $h^A
(C^A , \phi_c^A)$ span a 2-plane, i.e. a two-dimensional linear subspace of
${\cal
W}$. More explicitly, an unnormalized basis of this 2-plane is $h_1^A$, $h_2^A$
with
\begin{equation}
h_1^A \equiv h^A (C^A =1 , \phi_c^A =0) \ , \ h_2^A \equiv h^A \left(C^A = 1 ,
\phi_c^A = \frac{\pi}{2}\right) \, ; \label{eq:B2}
\end{equation}
a generic vector in the 2-plane being $h^A (\lambda^A) = \lambda_1^A \, h_1^A +
\lambda_2^A \, h_2^A$ with $\lambda_1^A = C^A \cos \phi_c^A$, $\lambda_2^A = C^A
\sin \phi_c^A$. Similarly, the two-parameter family of ``exact'' vectors can be
written as $h^X (\lambda^X) = \lambda_1^X \, h_1^X + \lambda_2^X \, h_2^X$ with
the same definitions as above with the label $A$ being changed into $X$.
Optimizing over the phases means finding the maximum over the $\lambda^A$ and
$\lambda^X$ of
\begin{equation}
\cos \theta_{AX} = \langle \widehat{h}^A , \widehat{h}^X \rangle \equiv
\frac{\langle \lambda_1^A h_1^A + \lambda_2^A h_2^A , \lambda_1^X h_1^X +
\lambda_2^X h_2^X \rangle}{\Vert \lambda_1^A h_1^A + \lambda_2^A h_2^A \Vert \,
\Vert \lambda_1^X h_1^X + \lambda_2^X h_2^X \Vert} \, , \label{eq:B3}
\end{equation}
where $\widehat{h}^A$ denotes the unit vector $h^A (\lambda) / \Vert h^A
(\lambda)\Vert$.

Directly attempting to maximize $\cos \theta_{AX} (\lambda^A , \lambda^X)$
is very
cumbersome. The problem is, however, reduced to an easy one if one introduces
orthonormalized bases in both 2-planes: say $(e_1^A , e_2^A)$ in the
$A$-(2-plane)
and $(e_1^X , e_2^X)$ in the $X$-one, with $\langle e_a^A , e_b^A \rangle =
\delta_{ab} = \langle e_a^X , e_b^X \rangle$; $a,b=1,2$. For instance, these
orthonormalized bases can be defined as
\begin{eqnarray}
e_1^A & \equiv  & \Vert h_1^A \Vert^{-1} \, h_1^A \, ,\label{eq:B4} \\
e_2^A & \equiv  & \Vert h_1^A \Vert^{-1} \, [\Vert h_1^A \Vert^2 \, \Vert h_2^A
\Vert^2 - \langle h_1^A ,h_2^A \rangle^2]^{-\frac{1}{2}} \, \{ \Vert h_1^A
\Vert^2
\, h_2^A - \langle h_1^A , h_2^A \rangle \, h_1^A \} \, , \nonumber
\end{eqnarray}
for the $A$-plane, and similarly for the $X$-plane.

The overlap (\ref{eq:B3}) is then the scalar product between two unit vectors (one
in each plane) which can be parametrized as $\cos \theta_{\alpha \beta} =
\langle
e_{\alpha}^A , e_{\beta}^X \rangle$ where $e_{\alpha}^A = \cos \alpha e_1^A
+ \sin
\alpha e_2^A$, $e_{\beta}^X = \cos \beta e_1^X + \sin \beta e_2^X$. Let $P_X$
denote the {\it orthogonal projector} onto the $X$-plane, and $p_{\alpha}$
denote
the orthogonal projection of $e_{\alpha}^A$, i.e. explicitly
\begin{equation}
p_{\alpha} = P_X (e_{\alpha}^A) = \langle e_{\alpha}^A , e_1^X \rangle \,
e_1^X +
\langle e_{\alpha}^A , e_2^X \rangle \, e_2^X \, . \label{eq:B5}
\end{equation}
The scalar product $\langle e_{\alpha}^A , e_{\beta}^X \rangle$ is equal to
$\langle p_{\alpha} , e_{\beta}^X \rangle$. It is maximized over $\beta$ when
$e_{\beta}^X$ is parallel to $p_{\alpha}$, in which case its value is the
norm of
$p_{\alpha}$. This shows that the maximum of $\cos \theta_{AX}$ is equal to the
maximum over $\alpha$ of the norm of $p_{\alpha}$:
\begin{equation}
(\cos \theta_{AX})_{\max} = \max_{\alpha} \, \Vert p_{\alpha} \Vert \, .
\label{eq:B6}
\end{equation}
On the other hand,
\begin{equation}
p_{\alpha} = \cos \alpha\ p_1 + \sin \alpha\ p_2 \, , \label{eq:B7}
\end{equation}
where
\begin{eqnarray}
p_1 & \equiv  & P_X (e_1^A) = \langle e_1^A , e_1^X \rangle \, e_1^X + \langle
e_1^A , e_2^X \rangle \, e_2^X \, ,\nonumber \\
p_2 & \equiv  & P_X (e_2^A) = \langle e_2^A , e_1^X \rangle \, e_1^X + \langle
e_2^A , e_2^X \rangle \, e_2^X \, . \label{eq:B8}
\end{eqnarray}

In geometrical terms, $p_{\alpha}$ describes, as $\alpha$ varies, and {\it
ellipse} in the $X$-plane (the projection of the circle $e_{\alpha} = \cos
\alpha\ e_1 + \sin \alpha\ e_2$) and the maximum projection onto the $X$-plane
corresponds
to the semi-major axis. The square $\Vert p_{\alpha} \Vert^2 = \langle
p_{\alpha}
, p_{\alpha} \rangle$ reads
\begin{equation}
\Vert p_{\alpha} \Vert^2 = A \, \cos^2 \alpha + B \, \sin^2 \alpha + 2C \, \cos
\alpha \, \sin \alpha \, , \label{eq:B9}
\end{equation}
where
\begin{eqnarray}
A & \equiv  & \Vert p_1 \Vert^2 = \langle e_1^A , e_1^X \rangle^2 + \langle
e_1^A
, e_2^X \rangle^2 \, ,\nonumber \\
B & \equiv  & \Vert p_2 \Vert^2 = \langle e_2^A , e_1^X \rangle^2 + \langle
e_2^A
, e_2^X \rangle^2 \, ,\nonumber \\
C & \equiv  & \langle p_1 , p_2 \rangle = \langle e_1^A , e_1^X \rangle \,
\langle e_2^A , e_1^X \rangle + \langle e_1^A , e_2^X \rangle \,
\langle e_2^A , e_2^X \rangle  \, . \label{eq:B10}
\end{eqnarray}

Maximizing over $\alpha$ is now easy (using $\cos^2 \alpha = (1+\cos
2\alpha)/2$,
$\sin^2 \alpha = (1-\cos 2\alpha)/2$, $2\sin \alpha \, \cos \alpha = \sin 2
\alpha$ and maximizing over $2\alpha$) and yields finally
\begin{equation}
(\cos \theta_{AX})_{\max} = \left[ \frac{A+B}{2} + \left[ \left(
\frac{A-B}{2}\right)^2 + C^2 \right]^{\frac{1}{2}} \right]^{\frac{1}{2}} \, .
\label{eq:B11}
\end{equation}
Inserting the definitions of the orthonormalized vectors Eq. (\ref{eq:B4}) into the
definitions Eq. (\ref{eq:B10}) of $A$, $B$ and $C$, can allow one to express $(\cos
\theta_{AX})_{\max}$ only in terms of various scalar products of the initial
vectors $h_1^A$, $h_2^A$, $h_1^X$, $h_2^X$. It is easily checked that the final
answer does not depend on the choice of basis in the $A$- and $X$-planes,
and can
(if wished) be expressed only in terms of the ``2-forms'' $\omega^A \equiv h_1^A
\wedge h_2^A$ and $\omega^X \equiv h_1^X \wedge h_2^X$ (and of the Euclidean
structure of ${\cal W}$).

The result Eq. (\ref{eq:B11}) gives the best possible overlap when
optimizing separately over the phases of the exact and
approximate signals. 
This gives the mathematical measure of the
closeness of the two wave forms. However, in practice we do not
have access to the phase of the exact signal. It might happen that
the latter phase, i.e. equivalently the angle $\beta,$ is not
optimum. Therefore, a physically more relevant measure 
of the closeness of the two wave forms
(especially for the purpose of detection) 
is obtained by first optimizing over $\alpha$ (the parameter
we can dial) and then considering that $\beta$ has the worst
possible value. In terms of the geometric reasoning given above
one finds that the worst possible case corresponds to the 
semi-minor axis of the ellipse given by Eq.~(\ref{eq:B9}), i.e.
\begin{equation}
\min_\beta \max_\alpha \left (\cos \theta_{AX} \right ) = 
\left[ \frac{A+B}{2} - \left[ \left( \frac{A-B}{2}\right)^2 + 
C^2 \right]^{\frac{1}{2}} \right]^{\frac{1}{2}} \, .
\label{eq:B12}
\end{equation}
In our simulations we considered both measures of the closeness of the
two signals. We use Eq. (\ref{eq:B11}) when we study the mathematical
convergence and we use Eq. (\ref{eq:B12}) when we are interested in the
detection. We shall refer to Eq. (\ref{eq:B11}) as the {\it best} 
overlap and Eq. (\ref{eq:B12}) as the {\it minimax} overlap.

\vglue 1cm

\clearpage
\begin {table} 
\caption {Location of the last stable circular orbit determined by
the $T$- and $P$-approximants in the test mass case. The $P$-approximants
predict the exact location at orders $v^4$ and beyond. At order $v^2$
the last stable orbit is not defined by $P$-approximants.}
\begin {tabular}{ccc}
$n$  & $x^{T_n}_{\rm lso}/x^X_{\rm lso}$ & $x^{P_n}_{\rm lso}/x^X_{\rm lso}$ \\
\tablerule
  &   \multicolumn{2}{c}{$x^X_{\rm lso} = 0.1667$}\\
\tablerule
2 & 3.0000 & --- \\
4 & 1.4415 & 1.0000 \\
6 & 1.1705 & 1.0000 \\

\end {tabular}\label {lso.tm}
\end {table}

\begin {table} 
\caption {Overlap integrals of a test mass
wave form whose Fourier transform
is computed using stationary phase approximation with the same wave form
but whose Fourier transform is computed using numerical fast Fourier transform.
$n$ stands for the order of the approximant with $X$ denoting the 
exact wave form.}
\begin {tabular}{cccc}
$n$  & A$_0$ & B$_0$ & C$_0$\\
\tablerule
     &   $\left <T_n^0,T_n^0 \right > $ 
     &   $\left <T_n^0,T_n^0 \right > $ 
     &   $\left <T_n^0,T_n^0 \right > $\\ 
\tablerule
 0 & 1.000 & 	 0.995 &	 0.967 \\ 
 2 & 0.999 & 	 0.994 &	 0.970 \\ 
 4 & 0.999 & 	 0.990 &	 0.955 \\ 
 6 & 0.999 & 	 0.986 &	 0.944 \\ 
 8 & 0.999 & 	 0.988 &	 0.945 \\ 
 10 & 0.999 & 	 0.987 &	 0.940 \\ 
 $X$ & 0.999 & 	 0.978 &	 0.918 \\ 

\end {tabular}\label {stationary.phase}
\end {table}

\begin {table} 
\caption {{\it Faithfulness} of the $T$- and $P$-approximants in the 
test mass case. Values quoted are the {\it minimax} overlaps
(cf. Eq.~(\protect \ref {eq:B12})) together with the best possible
overlaps (cf. Eq.~(\ref {eq:B11})) in parenthesis.}
\begin {tabular}{ccccccc}
$n$  & \multicolumn{2}{c}{A$_0$}
     & \multicolumn{2}{c}{B$_0$}
     & \multicolumn{2}{c}{C$_0$}\\
\tablerule
     &   $\left <T_n^0,X^0 \right > $ 
     &   $\left <P_n^0,X^0 \right > $ 
     &   $\left <T_n^0,X^0 \right > $ 
     &   $\left <P_n^0,X^0 \right > $ 
     &   $\left <T_n^0,X^0 \right > $ 
     &   $\left <P_n^0,X^0 \right > $ \\
\tablerule
      4 & 0.893 & 	0.868 & 	0.713 & 	0.833 & 	0.760 & 	0.964 \\[\up] 
   & (0.894) & 	 (0.868) & 	 (0.720) & 	 (0.838) & 	 (0.795) & 	 (0.984)\\ 
 5 & 0.545 & 	0.979 & 	0.338 & 	0.973 & 	0.496 & 	0.991 \\[\up] 
   & (0.545) & 	 (0.980) & 	 (0.343) & 	 (0.974) & 	 (0.514) & 	 (0.998)\\ 
 6 & 0.971 & 	0.996 & 	0.965 & 	0.988 & 	0.962 & 	0.994 \\[\up] 
   & (0.972) & 	 (0.996) & 	 (0.968) & 	 (0.990) & 	 (0.968) & 	 (0.999)\\ 
 7 & 0.989 & 	0.997 & 	0.858 & 	0.986 & 	0.891 & 	0.993 \\[\up] 
   & (0.989) & 	 (0.998) & 	 (0.862) & 	 (0.988) & 	 (0.924) & 	 (0.999)\\ 
 8 & 0.972 & 	0.998 & 	0.711 & 	0.989 & 	0.854 & 	0.992 \\[\up] 
   & (0.972) & 	 (0.999) & 	 (0.717) & 	 (0.992) & 	 (0.872) & 	 (0.999)\\ 
 9 & 0.995 & 	0.996 & 	0.928 & 	0.992 & 	0.933 & 	0.994 \\[\up] 
   & (0.995) & 	 (0.996) & 	 (0.933) & 	 (0.994) & 	 (0.957) & 	 (0.999)\\ 
10 & 0.988 & 	--- & 	0.849 & 	--- & 	0.903 & 	--- \\[\up] 
   & (0.988) & 	 (---) & 	 (0.854) & 	 (---) & 	 (0.937) & 	 (---)\\ 
11 & 0.988 & 	1.000 & 	0.858 & 	1.000 & 	0.908 & 	1.000 \\[\up] 
   & (0.989) & 	 (1.000) & 	 (0.862) & 	 (1.000) & 	 (0.940) & 	 (1.000)\\ 

\end {tabular}\label {faithfulness.tm}
\end {table}

\begin {table} 
\caption {Cauchy convergence of the $T$- and $P$-approximants 
in the test mass case. Values quoted are the best possible overlaps.}
\begin {tabular}{ccccccc}
$n$  & \multicolumn{2}{c}{A$_0$}
     & \multicolumn{2}{c}{B$_0$}
     & \multicolumn{2}{c}{C$_0$}\\
\tablerule
     &   $\left <T_n^0,T_{n+1}^0 \right > $ 
     &   $\left <P_n^0,P_{n+1}^0 \right > $ 
     &   $\left <T_n^0,T_{n+1}^0 \right > $ 
     &   $\left <P_n^0,P_{n+1}^0 \right > $ 
     &   $\left <T_n^0,T_{n+1}^0 \right > $ 
     &   $\left <P_n^0,P_{n+1}^0 \right > $ \\
\tablerule
      4 & 0.496 & 	0.918 & 	0.374 & 	0.914 & 	0.653 & 	0.992 \\[0pt]
 5 & 0.528 & 	0.984 & 	0.330 & 	0.986 & 	0.541 & 	0.999 \\[0pt]
 6 & 0.953 & 	1.000 & 	0.770 & 	1.000 & 	0.957 & 	1.000 \\[0pt]
 7 & 0.988 & 	0.999 & 	0.913 & 	0.997 & 	0.982 & 	0.998 \\[0pt]
 8 & 0.985 & 	0.999 & 	0.833 & 	0.999 & 	0.945 & 	0.999 \\[0pt]
 9 & 0.997 & 	0.999 & 	0.965 & 	0.994 & 	0.985 & 	0.999 \\[0pt]
10 & 1.000 & 	--- & 	0.997 & 	--- & 	0.999 & 	--- \\[0pt]

\end {tabular}\label {cauchy.tm}
\end {table}

\begin {table}
\caption {{\it Effectualness} of the $T$- and $P$-approximants 
in the test mass case. Values quoted are the {\it minimax} overlaps
together with the percentage bias in the estimation of total mass
$100 \left (1-m^A/m \right )$ in parenthesis.}
\begin {tabular}{ccccccc}
$n$  & \multicolumn{2}{c}{A$_0$}
     & \multicolumn{2}{c}{B$_0$}
     & \multicolumn{2}{c}{C$_0$}\\
\tablerule
     &   $\left < \left <T_n^0,X^0 \right > \right> $ 
     &   $\left < \left <P_n^0,X^0 \right > \right > $ 
     &   $\left < \left <T_n^0,X^0 \right > \right> $ 
     &   $\left < \left <P_n^0,X^0 \right > \right > $ 
     &   $\left < \left <T_n^0,X^0 \right > \right> $ 
     &   $\left < \left <P_n^0,X^0 \right > \right > $ \\
\tablerule
      4 &  $ 0.993 $ & 	 $ 0.973 $ & 	 $ 0.971 $ & 	 $ 0.999 $ & 	 $ 0.899 $ & 	 $ 0.998 $ \\[\up]& 
 ( $ 0.036 $) & 	 ( $ -0.079 $) & 	 ( $ -18.408 $) & 	 ( $ -3.361 $) & 	 ( $ -2.658 $) & 	 ( $ -0.649 $) \\[0pt]
 5 &  $ 0.824 $ & 	 $ 0.996 $ & 	 $ 0.986 $ & 	 $ 0.999 $ & 	 $ 0.806 $ & 	 $ 0.999 $ \\[\up]& 
 ( $ -0.382 $) & 	 ( $ -0.029 $) & 	 ( $ -104.846 $) & 	 ( $ -1.428 $) & 	 ( $ -15.040 $) & 	 ( $ -0.314 $) \\[0pt]
 6 &  $ 0.990 $ & 	 $ 1.000 $ & 	 $ 0.983 $ & 	 $ 0.997 $ & 	 $ 0.988 $ & 	 $ 0.999 $ \\[\up]& 
 ( $ 0.039 $) & 	 ( $ -0.004 $) & 	 ( $ -2.875 $) & 	 ( $ -0.040 $) & 	 ( $ -0.605 $) & 	 ( $ 0.000 $) \\[0pt]

\end {tabular}\label {effectualness.tm}
\end {table}

\begin {table}
\caption {Cauchy convergence of the $T$- and $P$-approximants 
in the comparable mass case. Values quoted are the best possible 
overlaps Eq.~(\protect\ref{eq:B11}).}
\begin {tabular}{ccccccc}
$n$  & \multicolumn{2}{c}{A}
     & \multicolumn{2}{c}{B}
     & \multicolumn{2}{c}{C}\\
\tablerule
     &   $\left <T_n^\eta,T_{n+1}^\eta \right > $ 
     &   $\left <P_n^\eta,P_{n+1}^\eta \right > $ 
     &   $\left <T_n^\eta,T_{n+1}^\eta \right > $ 
     &   $\left <P_n^\eta,P_{n+1}^\eta \right > $ 
     &   $\left <T_n^\eta,T_{n+1}^\eta \right > $ 
     &   $\left <P_n^\eta,P_{n+1}^\eta \right > $ \\
\tablerule
      3 & 0.904 & 	0.248 & 	0.482 & 	0.181 & 	0.703 & 	0.374 \\[0pt]
 4 & 0.411 & 	0.863 & 	0.338 & 	0.872 & 	0.545 & 	0.985 \\[0pt]
 5 & 0.536 & 	0.976 & 	0.321 & 	0.979 & 	0.537 & 	0.995 \\[0pt]
 6 & 0.946 & 	0.982 & 	0.736 & 	0.996 & 	0.928 & 	0.999 \\[0pt]

\end {tabular}\label {cauchy.fm}
\end {table}

{
\begin {table} 
\caption {Location of the last stable circular orbit determined by
the $T$- and $P$-approximants in the finite mass case for different
values of the parameter $\kappa_0$. At order $v^2$ the last stable 
orbit is not defined by $P$-approximants.  At orders $v^4$ and beyond 
the $P$-approximants predict the location of lso pretty well.}
\begin {tabular}{ccccc}
 & \multicolumn{2}{c}{$\eta=1/4$} & \multicolumn{2}{c}{$\eta=14/(11.4)^2$} \\
\tablerule
$n$  & $x^{T_n}_{\rm lso}/x^X_{\rm lso}$ & $x^{P_n}_{\rm lso}/x^X_{\rm lso}$ 
     & $x^{T_n}_{\rm lso}/x^X_{\rm lso}$ & $x^{P_n}_{\rm lso}/x^X_{\rm lso}$ \\
\tablerule
&    \multicolumn{4}{c}{$\kappa_0=-1$}   \\
\tablerule
         & \multicolumn{2}{c}{$x^X_{\rm lso}=0.1600$} 
    & \multicolumn{2}{c}{$x^X_{\rm lso}=0.1636$} \\
\tablerule
 2  &  2.8852 & ---    & 	 2.9495 & ---    \\
 4  &  1.5991 & 1.2415 & 	 1.5060 & 1.0976 \\
 6  &  1.2652 & 1.0205 & 	 1.2078 & 1.0035 \\

\tablerule
&    \multicolumn{4}{c}{$\kappa_0=47/39$} \\
\tablerule
         & \multicolumn{2}{c}{$x^X_{\rm lso}=0.1986$} 
    & \multicolumn{2}{c}{$x^X_{\rm lso}=0.1792$} \\
\tablerule
 2  &  2.3240 & ---    & 	 2.6934 & ---    \\
 4  &  1.2881 & 1.0000 & 	 1.3752 & 1.0023 \\
 6  &  1.1020 & 1.0000 & 	 1.1408 & 1.0000 \\

\tablerule
&    \multicolumn{4}{c}{$\kappa_0=2$}    \\
\tablerule
         & \multicolumn{2}{c}{$x^X_{\rm lso}=0.2145$} 
    & \multicolumn{2}{c}{$x^X_{\rm lso}=0.1852$}\\
\tablerule
 2  &  2.1517 & ---    & 	 2.6060 & ---    \\
 4  &  1.1925 & 0.9258 & 	 1.3306 & 0.9698 \\
 6  &  1.0568 & 1.0042 & 	 1.1186 & 1.0005 \\

\end {tabular}\label {lso.fm}
\end {table}
}

\begin {table}
\caption {Robustness of the $T$- and $P$-approximants in the 
comparable mass case: {\it Faithfulness}. Values quoted are 
the {\it minimax} overlaps together with the best possible 
overlaps, Eq.~(\protect\ref{eq:B11}), in parenthesis.  System 
D corresponds to a binary consisting of stars of masses 
$20~M_\odot$ and $1.4~M_\odot.$ In this extreme mass ratio case 
the $P$-approximants at $O(v^5)$ are NOT faithful (overlaps 
$<96.5\%$).}
\begin {tabular}{ccccccccc}
$n$  & \multicolumn{2}{c}{A}
     & \multicolumn{2}{c}{B}
     & \multicolumn{2}{c}{C}
     & \multicolumn{2}{c}{D}\\
\tablerule
     &   $\left <T_n^\eta,X^\eta \right > $
     &   $\left <P_n^\eta,X^\eta \right > $
     &   $\left <T_n^\eta,X^\eta \right > $
     &   $\left <P_n^\eta,X^\eta \right > $
     &   $\left <T_n^\eta,X^\eta \right > $
     &   $\left <P_n^\eta,X^\eta \right > $
     &   $\left <T_n^\eta,X^\eta \right > $
     &   $\left <P_n^\eta,X^\eta \right > $\\
\tablerule
&   \multicolumn{8}{c}{$\kappa_0=-1$}\\
\tablerule
      4 & 0.601 & 	0.875 & 	0.882 & 	0.735 & 	0.826 & 	0.825 & 	0.467 & 	0.696 \\[\up] 
   & (0.602) & 	 (0.875) & 	 (0.887) & 	 (0.742) & 	 (0.860) & 	 (0.841) & 	 (0.476) & 	 (0.705)\\ 
 5 & 0.528 & 	0.954 & 	0.323 & 	0.875 & 	0.416 & 	0.842 & 	0.272 & 	0.824 \\[\up] 
   & (0.528) & 	 (0.954) & 	 (0.328) & 	 (0.881) & 	 (0.432) & 	 (0.873) & 	 (0.279) & 	 (0.841)\\ 
 6 & 0.989 & 	0.987 & 	0.968 & 	0.990 & 	0.910 & 	0.991 & 	0.896 & 	0.998 \\[\up] 
   & (0.990) & 	 (0.988) & 	 (0.972) & 	 (0.992) & 	 (0.917) & 	 (0.996) & 	 (0.900) & 	 (0.999)\\ 

\tablerule
&   \multicolumn{8}{c}{$\kappa_0=47/39$}\\
\tablerule
      4 & 0.566 & 	0.810 & 	0.908 & 	0.793 & 	0.921 & 	0.954 & 	0.539 & 	0.800 \\[\up] 
   & (0.567) & 	 (0.810) & 	 (0.912) & 	 (0.795) & 	 (0.946) & 	 (0.976) & 	 (0.550) & 	 (0.813)\\ 
 5 & 0.553 & 	0.983 & 	0.335 & 	0.975 & 	0.490 & 	0.989 & 	0.283 & 	0.961 \\[\up] 
   & (0.554) & 	 (0.983) & 	 (0.339) & 	 (0.975) & 	 (0.509) & 	 (0.999) & 	 (0.290) & 	 (0.965)\\ 
 6 & 0.968 & 	0.998 & 	0.923 & 	0.997 & 	0.974 & 	0.997 & 	0.940 & 	0.990 \\[\up] 
   & (0.968) & 	 (0.998) & 	 (0.926) & 	 (0.999) & 	 (0.983) & 	 (0.999) & 	 (0.942) & 	 (0.996)\\ 

\tablerule
&   \multicolumn{8}{c}{$\kappa_0=2$}\\
\tablerule
      4 & 0.557 & 	0.779 & 	0.894 & 	0.813 & 	0.896 & 	0.985 & 	0.586 & 	0.843 \\[\up] 
   & (0.557) & 	 (0.780) & 	 (0.896) & 	 (0.817) & 	 (0.905) & 	 (0.989) & 	 (0.598) & 	 (0.848)\\ 
 5 & 0.563 & 	0.989 & 	0.339 & 	0.995 & 	0.507 & 	0.978 & 	0.286 & 	0.992 \\[\up] 
   & (0.563) & 	 (0.989) & 	 (0.344) & 	 (0.995) & 	 (0.531) & 	 (0.996) & 	 (0.294) & 	 (0.992)\\ 
 6 & 0.964 & 	0.999 & 	0.883 & 	0.978 & 	0.955 & 	0.997 & 	0.939 & 	0.984 \\[\up] 
   & (0.965) & 	 (0.999) & 	 (0.889) & 	 (0.978) & 	 (0.970) & 	 (0.998) & 	 (0.949) & 	 (0.992)\\ 

\end {tabular}\label {faithfulness.fm}
\end {table}
\clearpage

\begin {table}
\caption {Robustness of the $T$- and $P$-approximants in the comparable mass 
case: {\it Effectualness}.  Values quoted are the {\it minimax} overlaps
Eq.~(\protect \ref{eq:B12})
together with the percentage bias in the estimation of total mass
$100 \left (1-m^A/m \right )$ in parenthesis.}
\begin {tabular}{ccccccc}
$n$  & \multicolumn{2}{c}{A}
     & \multicolumn{2}{c}{B}
     & \multicolumn{2}{c}{C}\\
\tablerule
     &   $\left < \left <T_n^\eta,X^\eta \right > \right > $
     &   $\left < \left <P_n^\eta,X^\eta \right > \right > $
     &   $\left < \left <T_n^\eta,X^\eta \right > \right > $
     &   $\left < \left <P_n^\eta,X^\eta \right > \right > $
     &   $\left < \left <T_n^\eta,X^\eta \right > \right > $
     &   $\left < \left <P_n^\eta,X^\eta \right > \right > $\\
\tablerule
&   \multicolumn{6}{c}{$\kappa_0=-1$}\\
\tablerule
      4 &  $ 0.891 $ & 	 $ 0.970 $ & 	 $ 0.962 $ & 	 $ 0.997 $ & 	 $ 0.881 $ & 	 $ 0.940 $ \\[\up]& 
 ($ 0.268 $) & 	 ($ -0.089 $) & 	 ($ -12.550 $) & 	 ($ -7.582 $) & 	 ($ -2.217 $) & 	 ($ -2.779 $) \\[0pt]
 5 &  $ 0.811 $ & 	 $ 0.989 $ & 	 $ 0.713 $ & 	 $ 0.996 $ & 	 $ 0.664 $ & 	 $ 0.932 $ \\[\up]& 
 ($ -0.400 $) & 	 ($ -0.039 $) & 	 ($ -26.491 $) & 	 ($ -4.907 $) & 	 ($ -15.860 $) & 	 ($ -2.660 $) \\[0pt]
 6 &  $ 0.995 $ & 	 $ 0.991 $ & 	 $ 0.982 $ & 	 $ 0.999 $ & 	 $ 0.955 $ & 	 $ 0.998 $ \\[\up]& 
 ($ 0.011 $) & 	 ($ 0.025 $) & 	 ($ -0.869 $) & 	 ($ -0.104 $) & 	 ($ -0.999 $) & 	 ($ 0.080 $) \\[0pt]

\tablerule
&   \multicolumn{6}{c}{$\kappa_0=47/39$}\\
\tablerule
      4 & $ 0.868 $ & 	$ 0.965 $ & 	$ 0.975 $ & 	$ 0.999 $ & 	$ 0.972 $ & 	$ 0.997 $ \\[\up]& 
 ($ 0.214 $) & 	 ($ -0.111 $) & 	 ($ -6.961 $) & 	 ($ -3.618 $) & 	 ($ 1.406 $) & 	 ($ -1.524 $) \\[0pt]
 5 & $ 0.831 $ & 	$ 0.999 $ & 	$ 0.981 $ & 	$ 1.000 $ & 	$ 0.810 $ & 	$ 0.999 $ \\[\up]& 
 ($ -0.429 $) & 	 ($ -0.032 $) & 	 ($ -97.241 $) & 	 ($ -1.118 $) & 	 ($ -23.654 $) & 	 ($ -0.205 $) \\[0pt]
 6 & $ 0.984 $ & 	$ 0.998 $ & 	$ 0.989 $ & 	$ 0.999 $ & 	$ 0.996 $ & 	$ 1.000 $ \\[\up]& 
 ($ 0.025 $) & 	 ($ 0.000 $) & 	 ($ 1.004 $) & 	 ($ -0.157 $) & 	 ($ 0.391 $) & 	 ($ 0.019 $) \\[0pt]

\tablerule
&   \multicolumn{6}{c}{$\kappa_0=2$}\\
\tablerule
      4 & $ 0.863 $ & 	$ 0.961 $ & 	$ 0.968 $ & 	$ 0.998 $ & 	$ 0.996 $ & 	$ 0.997 $ \\[\up]& 
 ($ 0.357 $) & 	 ($ -0.121 $) & 	 ($ -1.300 $) & 	 ($ -2.874 $) & 	 ($ 1.699 $) & 	 ($ -0.450 $) \\[0pt]
 5 & $ 0.836 $ & 	$ 1.000 $ & 	$ 0.983 $ & 	$ 1.000 $ & 	$ 0.988 $ & 	$ 0.996 $ \\[\up]& 
 ($ -0.339 $) & 	 ($ -0.025 $) & 	 ($ -92.233 $) & 	 ($ -0.056 $) & 	 ($ -75.053 $) & 	 ($ 0.000 $) \\[0pt]
 6 & $ 0.981 $ & 	$ 0.999 $ & 	$ 0.993 $ & 	$ 0.995 $ & 	$ 0.998 $ & 	$ 1.000 $ \\[\up]& 
 ($ 0.032 $) & 	 ($ 0.000 $) & 	 ($ 0.688 $) & 	 ($ -0.708 $) & 	 ($ 1.106 $) & 	 ($ -0.154 $) \\[0pt]

\end {tabular}\label {effectualness.fm}
\end {table}
\clearpage

\clearpage

\clearpage
\begin {figure}[h]
\centerline {\epsfxsize 6 true in  \epsfbox {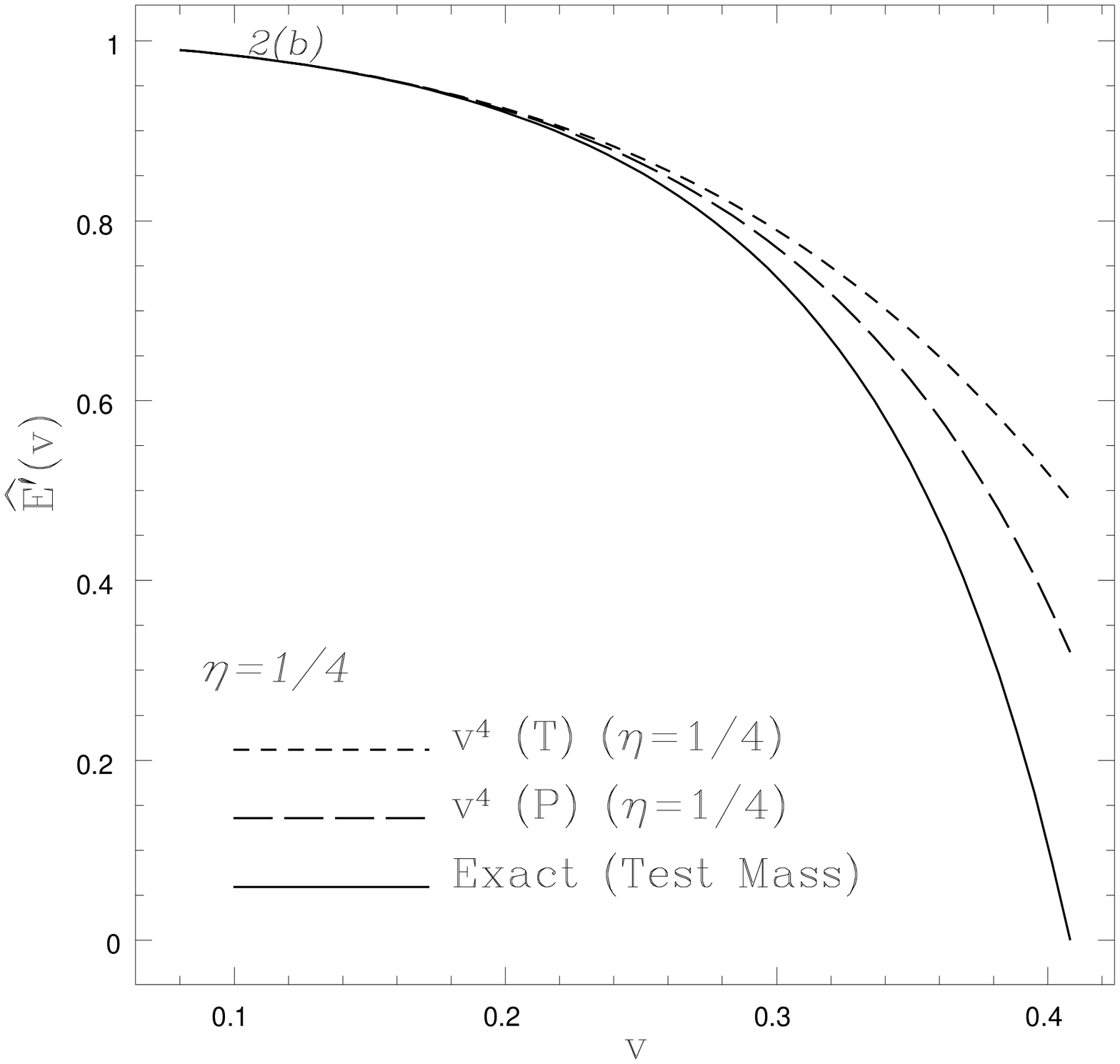} }
\end {figure}

\clearpage

\clearpage
\begin {figure}[t]
\centerline {\epsfxsize 6 true in  \epsfbox {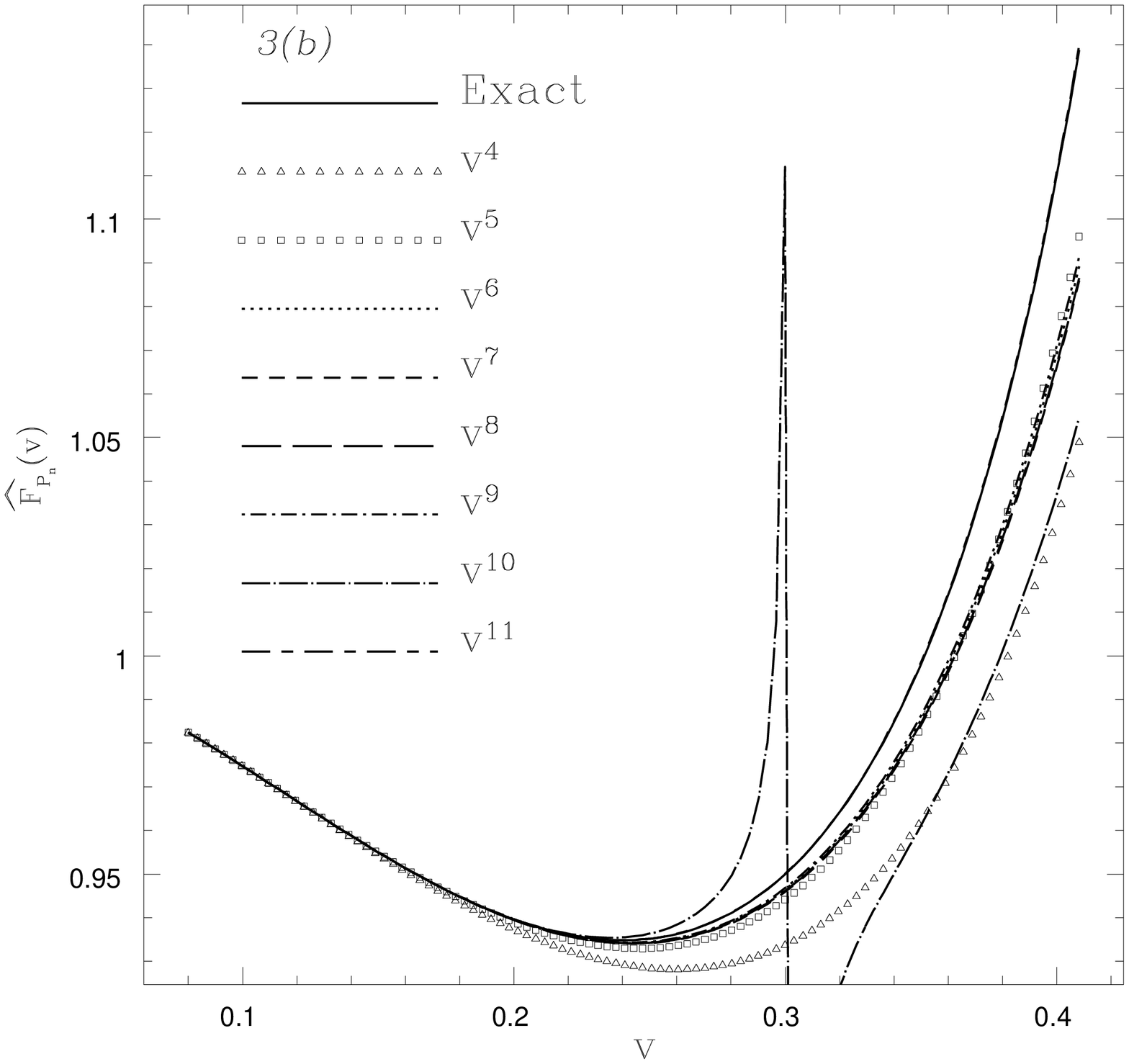} }
\end {figure}

\clearpage

\clearpage
\begin {figure}[t]
\centerline {\epsfxsize 6 true in  \epsfbox {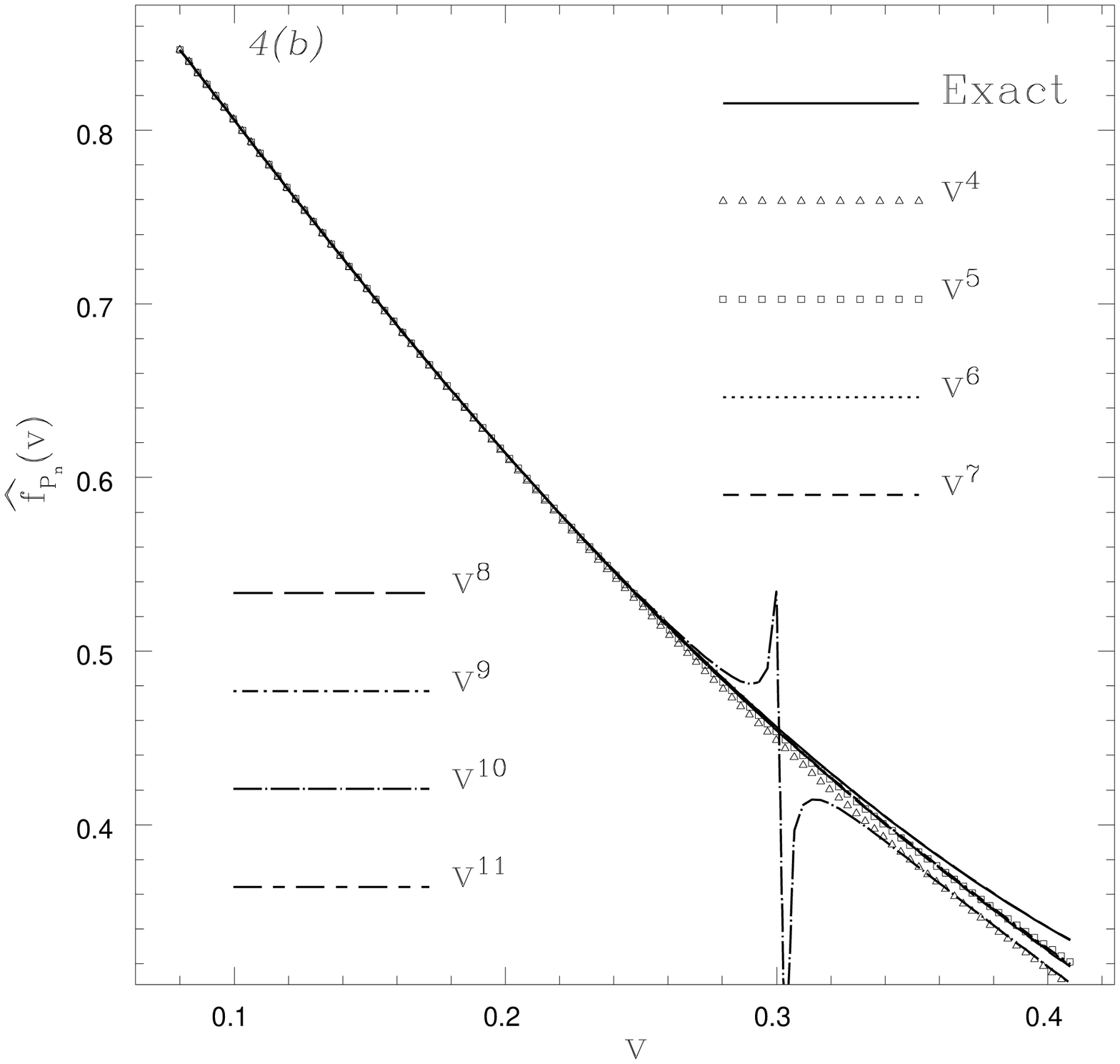} }
\end {figure}

\clearpage

\clearpage

\clearpage

\clearpage
\begin {figure}[t]
\centerline {\epsfxsize 6 true in  \epsfbox {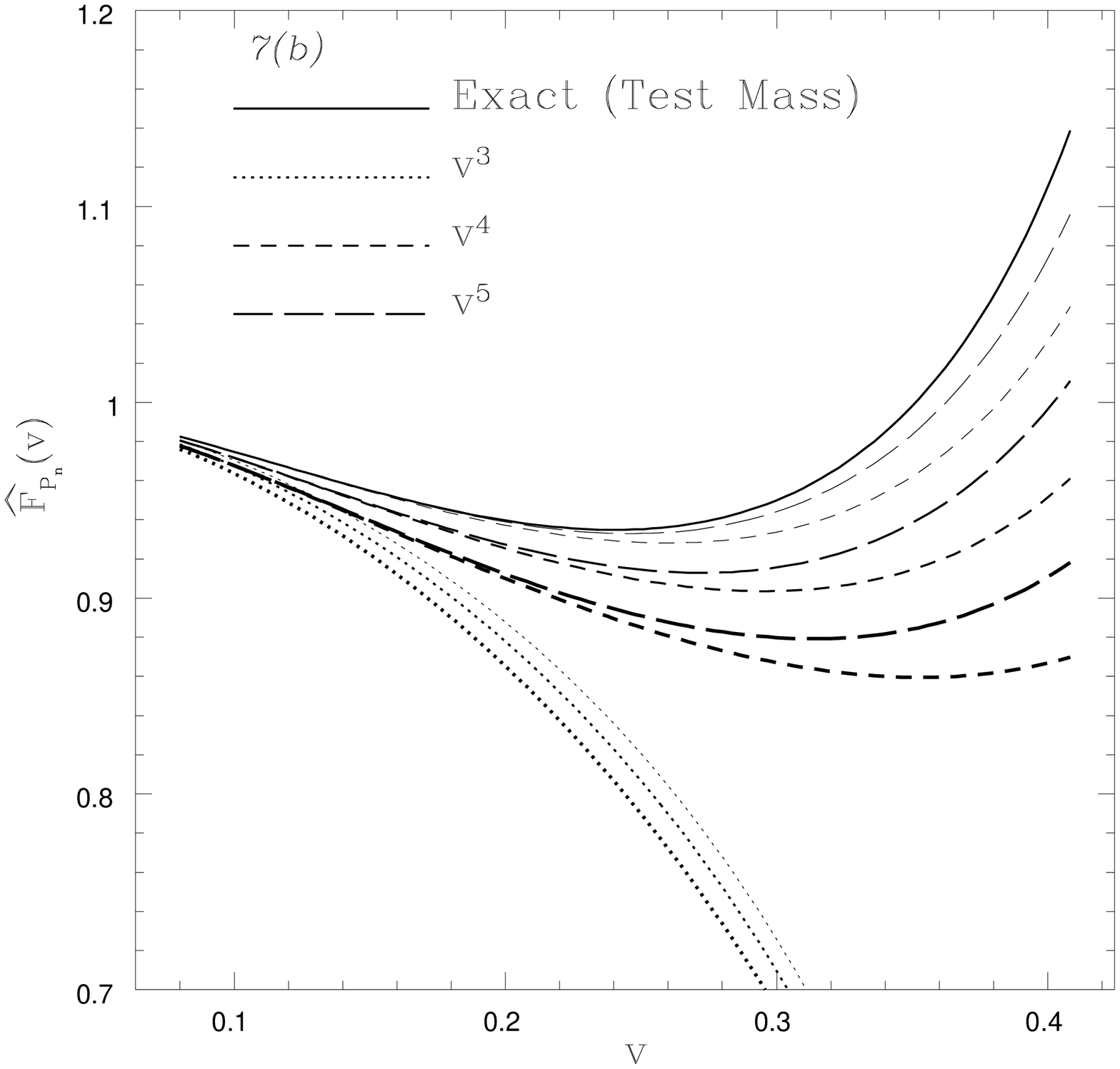} }
\end {figure}

\clearpage


\begin{thebibliography}{99}
\bibitem {ftn1}The most accuracy-demanding case is the
wave form emitted by a binary system made of a ($\sim 1.4 m_{\odot}$)
neutron star and a ($\gsim 10 m_{\odot}$) black hole \cite{TTS97,P97}.
\bibitem{ftn2}
In the present work we shall only consider as parameters of
the signal (besides an overall amplitude): two ``kinematical'' parameters
$t_c$, $\Phi_c$ (a reference time and a reference phase); and two
``dynamical'' ones $m_1$, $m_2$ or equivalently $m$, $\eta$. Our intention
here is to compare the performance of new approximants versus standard ones.
Evidently, our results can then be trivially generalized by constructing the
$P$-approximants of more general wave forms if the need arises.
\bibitem{ftn3}
We restrict our attention to circular(ized) systems.
\bibitem{ftn4}
We henceforth use units such that $G=c=1$.
\bibitem{ftn5}
For simplicity, we consider only diagonal approximants $(n,n)$.
\bibitem {caution:phase} Optimization over phase $\phi$ is quite 
subtle and it is treated at length in Appendix~\ref{appendix:phase}.
\bibitem{eventrate}
A loss in the overlap of 3\% is equal to a reduction in the potential
event rate by 10\% (see Sec.\ref{sec:ambiguity} and Ref.\cite{bjo96} for
details).
\bibitem{ftn6}
The thresholds one may wish to retain to define the smallness 
of the biases will evidently depend on the context.
\bibitem{ftn7}
But not sufficient, though one will, in practice, assume it to be so.
\bibitem{ftn9}
The choice of $v_{\rm lso}^A$ does not matter as any
change in this quantity can be reabsorbed in $t_c$ and $\Phi_c$. However,
our approximants to $E'_A (v)$ will have a zero (physically corresponding
to the last stable orbit) and it will be natural and convenient to choose
$v_{\rm lso}^A$ as the location of this zero.
\bibitem{ftn10}
In other words, we are assuming some minimum
``structural stability'' when turning on $\eta$.
\bibitem{ftn11}
Note that we consistently label
the successive approximants by the order in velocity. E.g. a $2PN$-accurate
object has the label 4.
\bibitem{ftn12}
It is to respect the
consistency of the linearized gravitational equations 
$(\partial_{\nu} \, T^{\mu \nu} =0)$ that we had to take 
explicitly into account the interaction
accelerating the masses on their orbits.
\bibitem{free} E.g., if one factorized out the log factors in the 
Flux function of Eq. (\ref{eq:N35}) as in Eq. (\ref{eq:n34}) 
\[
F(v;\eta=0) = \frac{32}{5} \, \eta^2 \, v^{10} \, \left[ 1+
\ln\frac{v}{v_{\rm lso}
(\eta)} \, \left( \sum_k \ell_k \, v^k \right) + \cdots \right] \, \left[ \sum_k
F_k \, v^k \right] \,\] 
then the 
corresponding coefficients $F_k$ are given by $ F_k=A_k\,, k=0\cdots 5\;;
F_k=A_k+B_k\ln v_{\rm lso}\,, k=6\cdots 11 $. The $\ell_k$ are as given by
Eq. (\ref{eq:n38})
\bibitem{ftn14}
As ${\cal A} (h,g)$ is the infinite-dimensional
generalization of the cosine of the angle between two unit vectors, we
could define
the angular distance between two normalized wave forms $(\Vert h \Vert = 1
= \Vert
g \Vert)$ by $\theta = \arccos \, {\cal A} (h,g)$ and the corresponding chord
by $\Vert h-g \Vert = 2 \sin  \frac{\theta}{2} = \sqrt{2(1-{\cal A}
(h,g))}$.
\bibitem {caution:p43}
Since our $P_4^3$-approximant of flux has a pole in the region
of interest we have instead used the $P_3^4$-approximant
in computing the overlaps and have not
used the $P^5_5$-approximant (which also has a pole) at all.
\end{thebibliography}
\end {document}